\documentclass[amsmath,amssymb,amsfonts,aps,pre,superscriptaddress,bibnotes,showpacs,showkeys,longbibliography,10pt,twocolumn]{revtex4-1}

\usepackage{epsfig}
\usepackage{ulem}
\usepackage[english]{babel}
\usepackage{soul}

\usepackage{subcaption}
\usepackage{color}
\usepackage{hyperref}

\begin{document}

\title{Three-component Bose-Einstein condensates\\ and wetting without walls}
\author{Joseph O. Indekeu}
\affiliation{Institute for  Theoretical Physics, KU Leuven, BE-3001 Leuven, Belgium}
\author{Nguyen Van Thu}
\affiliation{Department of Physics, Hanoi Pedagogical University 2, Hanoi 100000, Vietnam}
\author{Jonas Berx}
\affiliation{Niels Bohr International Academy, Niels Bohr Institute, University of Copenhagen, Blegdamsvej 17, 2100 Copenhagen, Denmark}

\date{\today}

\begin{abstract}
In previous work within Gross-Pitaevskii (GP) theory for ultracold gases wetting phase transitions were predicted for a phase-segregated two-component Bose-Einstein condensate (BEC) adsorbed at an optical wall. The wetting phase diagram was found to depend on intrinsic atomic parameters, being the masses and the scattering lengths, and on the extrinsic wall boundary condition. Here we study wetting transitions in GP theory without an optical wall in a setting with three phase-segregated BEC components instead of two. The boundary condition is removed by replacing the wall with the third component and treating the three phases on an equal footing. This leads to an unequivocal wetting phase diagram that depends only on intrinsic atomic parameters. It features first-order and critical wetting transitions, and prewetting phenomena. The phase boundaries are computed by numerical solution of the GP equations. In addition, useful analytic results are obtained by extending the established double-parabola approximation to three components. 
\end{abstract}

\maketitle 

\section{Introduction}
Ultracold gases provide an arena in which the laws of atomic quantum physics are at work in their theoretically most fundamental and experimentally most accessible manifestations \cite{Dalfovo,Pita}. Interatomic forces are tunable over many orders of magnitude in strength employing Feshbach resonances \cite{Inouye,Stan,Chin} and, at ultralow absolute temperature $T$, dilute gases display a panoply of cooperative effects \cite{Bloch, Gaunt, Navon}. A fascinating role herein is played by multi-component Bose-Einstein condensates (BEC), which can be manipulated directly and precisely at the atomic level to demonstrate surface and interface physics. This is different from the situation in classical ``thermal" fluid mixtures, in which thermodynamic fields and thermodynamic densities are controlled at a more macroscopic level. 

Among interfacial phenomena {\it wetting} is a very intriguing one \cite{deG}. The discovery of wetting phase transitions \cite{Cahn,ES,MC} provided a plethora of theoretical and experimental challenges \cite{deG,Dietrich,BonnRMP,BonnRoss}, phenomenologically connecting very diverse domains in surface and interfacial physics. In classical liquid mixtures, theoretically subtle and for a long time experimentally elusive {\it critical wetting} transitions were observed in 1996 \cite{Ragil} and 1999 \cite{RBM}. In type-I superconductors, the observation of a first-order interface delocalization (i.e., ``wetting") transition \cite{Kozhev} came about 12 years after its theoretical prediction \cite{IvL}. In BEC mixtures, wetting phase transitions were predicted in 2004 \cite{IVS}, but their experimental verification has to our knowledge not been accomplished hitherto. 

In 2004 first-order wetting phase transitions were predicted for a two-component BEC adsorbed at an optical hard wall \cite{IVS}. Subsequent extension of the theory, with more general wall boundary conditions, predicted a richer phase diagram with both first-order and critical wetting transitions \cite{VSIc}. Experimentally, wall boundary conditions can be realized using surface traps \cite{Rychtarik} with, ideally, square-well and flat-bottom confinement of the atoms \cite{Gaunt,Navon}. However, the use of a wall represents a drawback because theory predicts that the details of the boundary condition have an impact on the surface phase equilibria, which complicates experimental verification of the theoretical phase diagram. For example, in the wetting phase diagram predicted in \cite{VSIc} the order (first-order or critical) of the wetting transitions depends strongly on the “relative trap displacement”, a parameter not accessible in experiment. For a hard wall boundary condition, only first-order wetting transitions are predicted.

In order to obtain an unequivocal wetting phase diagram, in a space in which all variables are experimentally accessible, the optical wall is now omitted and replaced by a third BEC component that is treated on equal footing with the other two. This conceptual leap has been inspired by insights from wetting theory in phase-separated {\it classical} fluid mixtures.  Indeed, the density-functional theory for wetting in classical fluid three-phase equilibria is more satisfactory when the three phases are treated on an equal footing, instead of replacing one of the phases by a heuristic wall boundary condition \cite{RW}. The impact of this modification on the occurrence of wetting transitions can be drastic, in that nonwetting gaps can appear in the phase diagram \cite{IK,PR}. 

\begin{figure}[htp]
    \centering
    \includegraphics[width=0.9\linewidth]{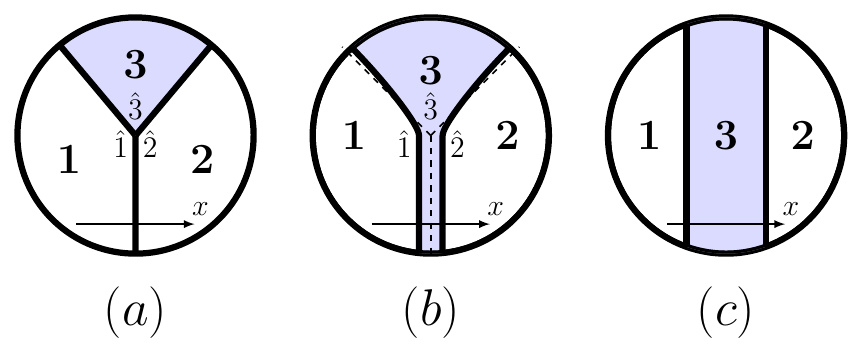}
    \caption{Theoretically predicted nonwet and wet three-component BEC configurations at three-phase coexistence. Shown are sketches, on a scale of typically 1 $\mu$m, of horizontal sections (in a disk-like trap) of the contact zone where three coexisting phases meet. (a) Nonwet: Condensates 1, 2 and 3 meet pairwise at their mutual interfaces, displaying dihedral angles $\hat 1$, $\hat 2$ and $\hat 3$ at a common line of contact. (b) Nonwet, with a thin film of 3 adsorbed at the 1-2 interface. (c) Wet: Contact angle $\hat 3$ is zero and a wetting layer of 3 intrudes between 1 and 2. In (a)-(c), the $x$-axis defines the direction of inhomogeneity along which the condensate wave functions vary in the calculation of the interfacial tensions.}
    \label{fig:trapconfig}
\end{figure}

It has been thoroughly demonstrated, theoretically \cite{Ho,Ao,sasaki,sasakiC,kobyakov2,takeuchi,maity,Naidon,ruban} and experimentally \cite{myatt,hall,miesner,stamper2,matthews,modugno,papp2,Thalhammer,tojo,mccarron,wang,Wacker,Grobner,burchianti,lee,Kwon,wilson}, that two-component BEC mixtures can display fascinating phase behavior and dynamical instabilities. Yet the new physics featured in BEC mixtures with more than two components has only recently spurred broad interest \cite{MaPe,Blom,Keiler,Roberts,JS,Edler,Saboo} and poses new experimental challenges.  Especially relevant for us is the GP theoretical study of interfacial phenomena in three-component BEC by Jimbo and Saito \cite{JS}. A third component, 3, adsorbed at the interface between condensates 1 and 2, can act as a surfactant and lower the 1-2 interfacial tension. Or, when the adsorbed layer is unstable, droplets of 3 form dynamically. Our present investigation of wetting phase transitions complements the study of surfactant behavior in \cite{JS}. 

In Fig.~\ref{fig:trapconfig} two-dimensional sections through characteristic nonwet and wet configurations for a three-component phase-segregated BEC at three-phase coexistence are depicted. In a nonwet state, three coexisting pure-component bulk phases and their mutual interfaces meet at a common line of contact. Condensates 1, 2 and 3 subtend the dihedral angles $\hat 1$, $\hat 2$ and $\hat 3$.  A simple criterion for wetting is ``Antonov's rule". For example, the 1-2 interface is {\bf nonwet} by 3 when the following inequality is strictly satisfied, and the 1-2 interface is {\bf wet} by 3 when the equality, aka Antonov's rule, holds \cite{RW},
\begin{equation}
\label{Antonov}
\gamma_{12(3)} \leq \gamma_{13} + \gamma_{23}.
\end{equation}
Here, $\gamma_{ij}$ is the $i$-$j$ interfacial tension in a BEC mixture consisting of two components $i$ and $j$ \cite{Ao}, and $\gamma_{12(3)} $ is the {\it three-component} 1-2 interfacial tension, allowing for the presence of a thin film of 3 adsorbed at the 1-2 interface. This film is stable (i.e., in equilibrium) if and only if its presence lowers the 1-2 interfacial tension, in which case 3 behaves as a surfactant \cite{JS}. If no such film of 3 is present at the 1-2 interface, then $\gamma_{12(3)} $ equals the $1$-$2$ interfacial tension $\gamma_{12}$.

This paper is organized as follows. Section II recalls the essentials of GP theory, defines the interfacial tensions and recapitulates the phenomenology of wetting phase transitions and applies this to the present context. Section III extends the double-parabola approximation to a mixture with three components. This approximation complements the numerical computations in GP theory with useful and insightful analytic approximations for interfacial tensions and wetting phase boundaries. 

In Section IV the wetting phase boundary is derived for mixtures in the weak-segregation regime, in GP theory (numerically precise) and in DPA (analytic approximation), and the results are shown to complement those obtained in \cite{JS} for the same regime. Section V is devoted to the strong segregation limit of condensated 1 and 2, and in that limit insightful analytic wetting phase diagrams are derived within DPA. In particular, an analytic approximation is obtained for a critical wetting phase boundary. In Section VI the GP equations are solved numerically for obtaining the wetting phase boundary in the more general, intermediate segregation regime. Evidence is obtained for the generic character of critical wetting transitions. 

In Section VII the wetting phase diagram off of three-phase coexistence is studied, focusing on prewetting phenomena and the nucleation transition for the wetting component 3, at two-phase coexistence of components 1 and 2. The validity of the mean-field theory at hand, and other issues relevant to experimental verification, such as the choices of atomic species and the trap configuration, are discussed in Section VIII. Section IX presents conclusions and an outlook on future work.

\section{GP theory, interfacial tensions and wetting in a three-component BEC mixture}
The GP theory applies the Bogoliubov mean-field theory for ultracold Bose gases to spatially inhomogeneous systems. A necessary condition for the validity of the mean-field approach is $\rho a^3 \ll 1$, with $\rho$ the number density and $a$ the s-wave scattering length \cite{Dalfovo,Pita}. The condition is satisfied for weakly interacting dilute gases, mixtures of which are considered in this paper. In the following the mean-field GP theory is adopted at $T=0$. This approach captures much of the physics of experimental interest in dilute Bose gases at ultralow $T$. For our purposes, the GP theory is cast as follows. The simple-harmonic-oscillator characteristic length of the conventional magnetic trap is assumed to be 5 $\mu$m or longer, which is large compared to the scattering lengths and healing lengths of the system. Therefore, in the calculations the confining potential is taken to be constant across the BEC interfaces of interest. 

In the grand canonical ensemble particle numbers are conveniently controlled by chemical potentials. Three components (atomic species) $i=1,2,3$, are assumed to be present in a volume $V$, with atomic masses $m_i$, chemical potentials $\mu_i$, macroscopic wave functions $\psi_i$ and (local) mean densities $n_i ({\bf r}) \equiv |\psi_i ({\bf r}) |^2 $. The wave functions play the roles of order parameters. Henceforth a notation is adopted to conform to that in \cite{VSIc}. The grand potential is the following functional,
\begin{equation}
\label{eq:Omega}
    \begin{split}
         \Omega [\{\psi_i\}] = &\sum_{i=1}^3   \int_V  d {\bf r} \, [\psi_i^*({\bf r})     \left [-\frac{\hbar^2}{2m_i}    \nabla^2 - \mu_i \right ] \psi_i({\bf r}) \\&+ \frac{G_{ii}}{2} |\psi_i({\bf r})|^4  ] 
         + \sum_{i < j} G_{ij} \int_V  d {\bf r} \,|\psi_i({\bf r})|^2 |\psi_j({\bf r})|^2 \;\\ \equiv &-\int_V \, d {\bf r} \; p({\bf r}),
 \end{split}
\end{equation}
with $-p({\bf r})$ the grand potential density. In the absence of flow, the phase of the complex wave function is constant, and one may choose the $\psi_i$ to be real-valued with $\psi_i \geq 0$.

Note that, for a homogeneous phase of a one-component BEC, the integrand is a constant, and the grand potential takes the value
\begin{equation}
    \Omega_{\rm bulk} = - P V,
\end{equation}
with $P$ the thermodynamic pressure in that phase in bulk. 
For a multicomponent BEC the chemical potentials are related to the pressures and densities in the bulk phases of each component, as follows. The equilibrium pressure $P_i$, and density $n_i$, for a pure and homogeneous
phase of species $i$ (in the absence of an external potential) are
\begin{eqnarray}
P_i &=& \frac{\mu_i^2}{2G_{ii}}\,, \\
n_i &\equiv &\psi_{i,\rm{bulk}}^2 = \frac{\mu_i}{G_{ii}}
\end{eqnarray}
and the healing length for component $i$ is 
\begin{equation}
    \xi_i = \frac{\hbar}{\sqrt{2m_i\mu_i}}\,.
\end{equation}
 
 The six interaction parameters $G_{ij}$ (with units of energy times volume) are linear in the six (a priori independent) scattering lengths $a_{ij}$,
\begin{equation}
    G_{ij} = 2\pi \hbar^2 a_{ij} (\frac{1}{m_i} + \frac{1}{m_j} ),\,\mbox{with} \;i,j = 1,2,3\,.
\end{equation}

The relative inter-species ($i \neq j$) interaction strength, or ``coupling", is
\begin{equation}
\label{eq:interaction_strength}
    K_{ij} \equiv G_{ij}/\sqrt{G_{ii}G_{jj}} = \frac{m_i+m_j}{2\sqrt{m_im_j}}\frac{a_{ij}}{\sqrt{a_{ii}a_{jj}}}\,.
\end{equation}
For sufficiently repulsive couplings, $K_{ij} > 1$, condensates $i$ and $j$ demix and phase segregate \cite{Ao, Roberts} and the completely immiscible case is considered for which all interspecies couplings exceed unity (cf. ${\cal E}_3^{im}$ in Fig.~1 of~\cite{Roberts}). Importantly, the couplings can be manipulated experimentally by tuning a scattering length. Using magnetic Feshbach resonance, any one of the scattering lengths, for example the interspecies $a_{ij}$, can be varied over several orders of magnitude~\cite{Inouye,Stan,Chin}.  

Henceforth two-phase equilibrium of condensates 1 and 2, $P_1 = P_2 \equiv  P$, is presupposed, so that a stable 1-2 interface exists. Condensate 3 is either metastable in bulk, $P_3 < P$, or coexists with 1 and 2 in a three-phase equilibrium, $P_3 = P$. The latter permits the study of wetting transitions, while the former is suitable for investigating prewetting phenomena \cite{IVS,VSIc}. Note that, at two-phase coexistence of condensates $i$ and $j$, their healing length ratio depends on atomic parameters alone, $\xi_i /\xi_j = (m_j \,a_{jj}/m_i\,a_{ii} )^{1/4}$.

The following spatial configuration is adopted. Condensates 1 and 2 are imposed as the bulk phases at $x \rightarrow - \infty$ and $x \rightarrow \infty$, respectively. The candidate wetting phase is condensate 3. It is convenient to define the auxiliary chemical potential value $\bar\mu_3$ and the auxiliary density value $\bar n_3$, which, respectively, $\mu_3$ and $n_3$ take when $P_3$ is tuned so as to equal $P$. So, 
\begin{eqnarray}
\bar \mu_3 &\equiv& \sqrt{\frac{G_{33}}{G_{11}}} \,\mu_1 = \sqrt{\frac{G_{33}}{G_{22}}}\,\mu_2\,, \\
\bar n_3 &\equiv& \frac{\bar \mu_3}{G_{33}}. 
\end{eqnarray}
One has $\mu_3 \leq \bar \mu_3 $ and $n_3 \leq \bar n_3$. The corresponding value of $\xi_3$ is denoted by $\bar \xi_3$,
\begin{equation}
  \bar \xi_3 \equiv  \frac{\hbar}{\sqrt{2m_3\bar \mu_3}}.
\end{equation} 

Note that, with these definitions, one has
\begin{equation}
 P_i = \frac{\mu_i n_i}{2}
\end{equation}
and, at coexistence of 1 and 2,
\begin{equation}
 P \equiv P_1 = P_2 \geq P_3 = \left (\frac{\mu_3}{\bar \mu_3}\right )^2P.
\end{equation} 

If rescalings are performed analogous to those described in \cite{VSIc},  $\psi_i \equiv \sqrt{n_i}\, \tilde \psi_i$, $i=1,2$ and $ \psi_3 \equiv \sqrt{\bar n_3} \,\tilde \psi_3$, $x  \equiv \xi_2 \,\tilde x$, one arrives at the three coupled GP ``equations of motion", with $j\in \{1,2,3\}$,
\begin{eqnarray}
\label{coupledGP}
   \left (\frac{\xi_1}{\xi_2}\right )^2    \frac{d^2\tilde \psi_1}{d\tilde x ^2}  &=& - \tilde  \psi_1 + \tilde \psi_1^3 + \Sigma_{j \neq 1} \,K_{1j} \,\tilde \psi_j^2 \,\tilde \psi_1, 
    \nonumber \\
    \frac{d^2\tilde \psi_2}{d\tilde x ^2} &=& - \tilde  \psi_2 + \tilde \psi_2^3 + \Sigma_{j \neq 2} \,K_{2j}\, \tilde \psi_j^2 \,\tilde \psi_2, 
    \nonumber \\
    \left (\frac{\bar\xi_3}{\xi_2}\right )^2    \frac{d^2\tilde \psi_3}{d\tilde x ^2} &=& - \frac{\mu_3}{\bar \mu_3}\tilde  \psi_3 + \tilde \psi_3^3 + \Sigma_{j \neq 3} \,K_{3j} \,\tilde \psi_j^2 \,\tilde \psi_3,\, 
\end{eqnarray}
with, of course, $K_{ij} \equiv K_{ji}$.
Note that the bulk values of $\tilde \psi_i$, $i=1,2$, are unity, whereas the bulk value of $\tilde \psi_3$ can be less than unity depending on the deviation from three-phase coexistence. Note also that the ratio $\bar\xi_3/\xi_2 = \sqrt{m_2  \mu_2/(m_3 \bar\mu_3)}$ is independent of $\mu_3$ and therefore can be kept constant when one goes off of three-phase coexistence by lowering $\mu_3$ below $\bar \mu_3$. The boundary conditions in bulk are
\begin{eqnarray}
    \tilde \psi_1 \rightarrow 1, &\; &\tilde \psi_{j\neq 1} \rightarrow 0, \;\; \mbox{for} \; \tilde x \rightarrow - \infty, \\
    \tilde \psi_2 \rightarrow 1, &\;& \tilde \psi_{j\neq 2} \rightarrow 0, \;\; \mbox{for} \; \tilde x \rightarrow  \infty\,.
\end{eqnarray}

A useful first integral or ``constant of the motion"  is obtained by multiplying the $i$-th GP equation of \eqref{coupledGP} by $d\tilde\psi_i/d\tilde x$, integrating it from $\tilde x =-\infty$ to an arbitrary point $\tilde x$, and summing up the three equations. It reads
\begin{eqnarray}
\label{first integral}
    &&   \left (\frac{\xi_1}{\xi_2}\frac{d\tilde \psi_1}{d\tilde x }\right )^2  + \left (\frac{d\tilde \psi_2}{d\tilde x }\right )^2  +    \left ( \frac{\bar\xi_3}{\xi_2}\frac{d\tilde \psi_3}{d\tilde x }\right )^2 \nonumber \\ &+&\tilde  \psi_1^2 +\tilde  \psi_2^2 +\frac{\mu_3}{\bar \mu_3} \tilde  \psi_3^2 - \frac{\tilde \psi_1^4}{2} - \frac{\tilde \psi_2^4}{2} - \frac{\tilde \psi_3^4}{2}\nonumber \\ &-& \sum_{i=1}^{3}\,\sum_{j > i} \,K_{ij} \,\tilde \psi_i^2 \,\tilde \psi_j^2 - \frac{1}{2} \equiv E_{\rm kin} + V = 0.
\end{eqnarray}
Note that the integration constant $1/2$ is determined by the boundary conditions in bulk, at $ \tilde x = \pm \infty$. The classical ``mechanical" interpretation of \eqref{first integral} is energy conservation in the form $E_{\rm kin}+V=0$, with $E_{\rm kin}$ the ``kinetic energy" (the three gradient squared terms) and $V$ the ``potential" (the remaining terms). Note that {\it minus} $V$ is a triple-well potential that represents the potential energy of the condensates. In the mechanical analogy of our thermodynamic problem a particle with three spatial coordinates, mimicked by the three $\tilde \psi_i$, moves from top to top in the triple-hill potential $V$ according to Newton's equation of motion, with $\tilde x$ playing the role of time \cite{RW}.

To calculate the interfacial tensions it suffices to consider a one-dimensional inhomogeneity, say along $x$, and to assume translational invariance along $y$ and $z$. Even in the nonwet state, when three interfaces meet at a common contact line with nontrivial dihedral angles, the three relevant interfacial tensions are calculated ``infinitely" far from the contact line, where the interfaces are simply {\it planar} surfaces between two bulk phases, with or without {\it planar} surfactant film of the third component. 

The interfacial tension is the {\it excess} grand potential {\it per unit area} of the inhomogeneous state that arises when the bulk states are fixed to be two different components. That is, for a one-dimensional inhomogeneity,
\begin{equation}
\gamma \equiv \int _{-\infty }^{\infty} d\, x \left (P-p(x)\right )\,.
\end{equation}
Note that the counterterm $P$ ensures that the integrand vanishes in the coexisting bulk phases at $x = \pm\infty$, so that $\gamma$ is indeed a finite ``excess" quantity, which picks up contributions in the interfacial zone and not in the bulk.

Virtually exact expressions have been derived for two-component $\gamma_{ij}$ \cite{Bar,VS}. For our purposes, high-precision numerical integrations provide $\gamma_{12(3)}$ as well as the $\gamma_{ij}$. For example, for component 1 at $x = -\infty$, which is indicated with ``$-\infty,1$" in the lower integration limit, and component 2 at $x = \infty$, indicated with ``$\infty,2$", with no surfactant present ($\tilde \psi_3=0$), and using the 
first integral \eqref{first integral} to eliminate the potential terms in the grand potential density \eqref{eq:Omega}, one arrives at the integral
\begin{equation}
\label{gamma12}
    \gamma_{12} \equiv 4 P \,\xi_2 \int _{-\infty,1 }^{\infty,2} d \tilde x  \;\left \{\left (\frac{\xi_1}{\xi_2}\frac{d\tilde \psi_1}{d\tilde x }\right )^2  + \left (\frac{d\tilde \psi_2}{d\tilde x }\right )^2       \right \},
\end{equation}
with $P_1 = P_2 = P$.

Likewise,
\begin{equation}
\label{gamma13}
    \gamma_{13} \equiv 4 P \,\xi_2 \int _{-\infty,1 }^{\infty,3} d \tilde x  \;\left \{\left (\frac{\xi_1}{\xi_2}\frac{d\tilde \psi_1}{d\tilde x }\right )^2 + \left ( \frac{\bar\xi_3}{\xi_2}\frac{d\tilde \psi_3}{d\tilde x }\right )^2       \right \},
\end{equation}
which is to be used only when 3 coexists with 1 and 2, that is, $P_3 = P$. And, under the same conditions of three-phase coexistence ($P_3=P$), 
\begin{equation}
\label{gamma23}
    \gamma_{23} \equiv 4 P \,\xi_2 \int _{-\infty,2 }^{\infty,3} d \tilde x  \;   \left \{\left (\frac{d\tilde \psi_2}{d\tilde x }\right )^2   + \left ( \frac{\bar\xi_3}{\xi_2}\frac{d\tilde \psi_3}{d\tilde x }\right )^2   \right \}\,.
\end{equation}
A similar calculation provides $\gamma_{12(3)}$,  
\begin{eqnarray}
\label{gamma123}
    \gamma_{12(3)} &\equiv  &4 P \,\xi_2 \int _{-\infty,1 }^{\infty,2} d \tilde x  \;\left \{\left (\frac{\xi_1}{\xi_2}\frac{d\tilde \psi_1}{d\tilde x }\right )^2  
    \right. \nonumber \\ &+ &\left.\left (\frac{d\tilde \psi_2}{d\tilde x }\right )^2  +    \left ( \frac{\bar\xi_3}{\xi_2}\frac{d\tilde \psi_3}{d\tilde x }\right )^2\right \},
\end{eqnarray}
which can be used at, and also off of three-phase coexistence, for $P_3 \leq  P$.

To investigate conceptually a transition in which the 1-2 interface is wet by 3, one starts by considering a nonwet state. Suppose the equilibrium 1-2 interface has no adsorbed film of 3. Its interfacial tension then equals $\gamma_{12}$ and is assumed to be higher than either $\gamma_{13}$ or $\gamma_{23}$, but lower than their sum, so $\gamma_{12} < \gamma_{13} + \gamma_{23}$. 
Previous experience with GP theory of wetting in BEC \cite{IVS} then suggests that, when $K_{13}$ and/or $K_{23}$ are decreased (towards unity), thereby lowering $\gamma_{13}$ and/or $\gamma_{23}$ (towards zero), while keeping $K_{12}$ constant, one may reach a state in which the following equality is realized: $\gamma_{12} = \gamma_{13} + \gamma_{23}$. This might signify that a phase transition takes place from the nonwet state (without any film of 3) to a 1-2 interface wet by a macroscopic layer of 3, and if so, it would typically be a wetting transition of first order (with metastable state continuations in mean-field theory). We will refer to this scenario as {\it strongly first-order} wetting.

However, a more subtle scenario is also possible in which the wetting transition is ``postponed". An equilibrium thin film of 3 may form at the nonwet 1-2 interface, the thickness of which increases as $K_{13}$ and/or $K_{23}$ are decreased. This is the case when the 1-2 interfacial tension decreases in the presence of a film of 3. Component 3 can then be regarded as a surfactant, and  $\gamma_{12(3)} < \gamma_{12}$. This possibility was clearly demonstrated and its mechanism explained in \cite{JS}. Consequently $K_{13}$ and/or $K_{23}$ must be {\it further} lowered in order to satisfy the condition for wetting, $\gamma_{12(3)} = \gamma_{13} + \gamma_{23}$. 

When this condition is reached a first-order wetting transition might take place between a nonwet state with a surfactant film of 3 and the wet state with a macroscopic wetting layer of 3.  However, another challenging possibility is a continuous or ``critical" wetting transition, in which the thickness of the surfactant film in the nonwet state increases continuously, and apparently without bound, upon reaching the condition for wetting. Note that in GP theory for a two-component BEC adsorbed at an optical wall, first-order wetting is predicted for a hard wall \cite{IVS}, while both  first-order and critical wetting are possibilities predicted for a softer wall \cite{VSIc}. 

\section{The double-parabola approximation for three components}
Complementary to numerical solution of the GP equations, one can also capture much of the physics of our problem by means of an analytic calculation, which is an extension to three components of the double-parabola approximation (DPA). The DPA has proven to be quite reliable for studying wetting in two-component GP theory (see, e.g., Fig.5 in \cite{Istatic}). The two-component DPA has provided useful analytic approximations to the interfacial tensions, derived in \cite{Istatic} and recalled here, using the labels $i,j \in \{1,2,3\}$ appropriate for the three-component mixture under study. These are,
\begin{eqnarray}
\label{sigmaDPA}
\gamma_{ij}^{(\rm DPA)} &=& 2\sqrt{2} \frac{\sqrt{(K_{ij}-1)/2}}{1+ \sqrt{(K_{ij}-1)/2}}P (\xi_i + \xi_j),
\end{eqnarray}
with $P$ the pressure at two-phase coexistence of components $i$ and $j$, and $i \neq j$. Note that \eqref{sigmaDPA} provides the DPA for the interfacial tensions \eqref{gamma12}, \eqref{gamma13} and \eqref{gamma23}.

Here an extension of the DPA is proposed, adapted to the three-component GP theory.  It consists of defining a piecewise harmonic approximation to the potential $V$ defined in \eqref{first integral} and solving piecewise linear GP equations in three adjacent domains, along the $\tilde x$-direction, associated with the three potential hills embodied in $V$. The nonlinear nature of the theory remains present through weak singularities at the domain junctions $\tilde x^- (\leq 0)$ and $\tilde x^+ (\geq 0)$, where wave functions and their first derivatives are continuous but their higher derivatives are not. The equilibrium wetting layer thickness, which is commonly considered as the (surface) order parameter associated with wetting, is $\tilde L \equiv \tilde x^+ - \tilde x^-$. 

In a first step one derives $V^{(\rm DPA)}$, the DPA for the potential $V$. In the leftmost domain I ($\tilde x < \tilde x^-$), $V$ is expanded to second order in the deviations from the bulk density for $\tilde\psi_1$, which equals unity, and in the deviations from zero for $\tilde\psi_2$ and $\tilde\psi_3$. In the middle domain ($\tilde x^-  < \tilde x < \tilde x^+ $)  the potential is expanded to second order in the deviations from its local maximum, about $\tilde\psi_3 = \sqrt{\mu_3/\bar \mu_3} = \sqrt{n_3/\bar n_3}$, and in the deviations from zero for $\tilde\psi_1$ and $\tilde\psi_2$. Note that at this point $V(\tilde\psi_1,\tilde\psi_2,\tilde\psi_3)$ takes the value
\begin{equation}
V(0,0,\sqrt{\frac{\mu_3}{\bar \mu_3}}) =  \frac{1}{2}\left [
\frac{\mu_3^2}{\bar \mu_3^2}-1 \right ]\leq 0,
\end{equation}
which indeed corresponds to a (local) maximum provided condensate 3 is sufficiently close to coexistence with 1 and 2, i.e., provided $1 \geq \mu_3/\bar \mu_3 > 1/K_{13}$ and $1 \geq \mu_3/\bar \mu_3 > 1/K_{23}$. In the rightmost domain III ($\tilde x > \tilde x^+$) labels 1 and 2 must be interchanged relative to domain I. 

For domain I  this leads to 
\begin{equation}
\label{ternaryDPAforVdomainI}
    V^{(\rm DPA)}_{\rm I} =
    -2 (1-\tilde \psi_1)^2  -(K_{12}-1) \tilde \psi_2^2 -(K_{13} - \frac{\mu_3}{\bar \mu_3})\tilde \psi_3^2. 
\end{equation}
In domain II  one gets
\begin{eqnarray}
\label{ternaryDPAforVdomainII}
    V^{(\rm DPA)}_{\rm II} &=
    -\frac{1}{2} + \frac{1}{2}\left (\frac{\mu_3}{\bar \mu_3}\right )^2
    -2 \frac{\mu_3}{\bar \mu_3}\left(\sqrt{\frac{\mu_3}{\bar \mu_3}}-\tilde \psi_3\right)^2  \nonumber \\&-(\frac{\mu_3}{\bar \mu_3}K_{13}-1) \tilde \psi_1^2 
    -(\frac{\mu_3}{\bar \mu_3}K_{23} - 1)\tilde \psi_2^2, 
\end{eqnarray}
and for domain III  the result is
\begin{equation}
\label{ternaryDPAforVdomainIII}
    V^{(\rm DPA)}_{\rm III} =
    -2 (1-\tilde \psi_2)^2  -(K_{12}-1) \tilde \psi_1^2 -(K_{23} - \frac{\mu_3}{\bar \mu_3})\tilde \psi_3^2. 
\end{equation}

In a second step one writes down the ensuing DPA for the GP equations in domains I, II and III, together with the analytic solutions for the wave functions in each domain. 
Whenever the second derivatives have a definite sign in a domain, this is indicated in the equations. The DPA equations of motion take the form, for domain I,
\begin{eqnarray}
\label{domainIeq1}
\left (\frac{\xi_1}{\xi_2}\right )^2    \frac{d^2\tilde \psi_1}{d\tilde x ^2}  &=& -2(1- \tilde  \psi_1) \leq 0, 
   \\ \label{domainIeq2} 
           \frac{d^2\tilde \psi_2}{d\tilde x ^2}  &=& (K_{12} -1) \tilde  \psi_2 \geq 0,  
      \\ \label{domainIeq3} 
  \left (\frac{\bar\xi_3}{\xi_2}\right )^2   \frac{d^2\tilde \psi_3}{d\tilde x ^2} &=& (K_{13}-\frac{\mu_3}{\bar \mu_3}) \tilde  \psi_3 \geq 0. 
\end{eqnarray}
The solutions in domain I read,
\label{domainIsolutions}
\begin{eqnarray}
\tilde\psi_1  &=&  1 - A_1 \,\exp{(\sqrt{2} \frac{\xi_2}{\xi_1}\tilde x)},
    \\
    \tilde \psi_2  &=&   A_2 \,\exp{(\sqrt{K_{12}-1} \,\tilde x)},
        \\
  \tilde \psi_3 &=& A_3 \,\exp{ (\sqrt{K_{13}-\frac{\mu_3}{\bar \mu_3}} \frac{\xi_2}{\bar\xi_3}\tilde x )}. 
   \end{eqnarray}

For domain II the DPA equations take the form,
\begin{eqnarray}
\label{domainIIeq1}
\left (\frac{\xi_1}{\xi_2}\right )^2    \frac{d^2\tilde \psi_1}{d\tilde x ^2}  &=& (\frac{\mu_3}{\bar \mu_3}K_{13}-1) \tilde  \psi_1, 
    \\ \label{domainIIeq2}
        \frac{d^2\tilde \psi_2}{d\tilde x ^2}  &=& (\frac{\mu_3}{\bar \mu_3} K_{23}-1) \tilde  \psi_2,      
       \\ \label{domainIIeq3}
  \left (\frac{\bar\xi_3}{\xi_2}\right )^2   \frac{d^2\tilde \psi_3}
{d\tilde x ^2} &=& -2  \,\frac{\mu_3}{\bar \mu_3}( \sqrt{\frac{\mu_3}{\bar \mu_3}} - \tilde  \psi_3) \leq 0. 
   \end{eqnarray}
The solutions in domain II read, 
\begin{eqnarray}
\label{domainIIsolutions1}
\tilde \psi_1  &=&    B_1 \,\exp{ (\sqrt{\frac{\mu_3}{\bar \mu_3}K_{13}-1} \,\frac{\xi_2}{\xi_1}\tilde x )} \nonumber\\&+& C_1 \exp{ (-\sqrt{\frac{\mu_3}{\bar \mu_3}K_{13}-1}\,\frac{\xi_2}{\xi_1}\tilde x )}, 
    \\
\label{domainIIsolutions2}
\tilde \psi_2  &=&    B_2 \,\exp{( \sqrt{\frac{\mu_3}{\bar \mu_3}K_{23}-1} \,\tilde x )} \nonumber\\&+& C_2 \exp{( -\sqrt{\frac{\mu_3}{\bar \mu_3}K_{23}-1}\,\tilde x )}, 
        \\
  \tilde \psi_3 &=& \sqrt{\frac{\mu_3}{\bar \mu_3}} + B_3 \,\exp{(\sqrt{2} \sqrt{\frac{\mu_3}{\bar \mu_3}} \frac{\xi_2}{\bar\xi_3}\tilde x) }\nonumber\\&+& C_3\, \exp{(-\sqrt{2} \sqrt{\frac{\mu_3}{\bar \mu_3}} \frac{\xi_2}{\bar\xi_3}\tilde x) }.
   \end{eqnarray}
   
For domain III  the DPA equations take the form,
\begin{eqnarray}
\label{domainIIIeq1}
\left (\frac{\xi_1}{\xi_2}\right )^2   \frac{d^2\tilde \psi_1}{d\tilde x ^2} &=& (K_{12} - 1) \tilde \psi_1 \geq 0, 
    \\ \label{domainIIIeq2}
    \frac{d^2\tilde \psi_2}{d\tilde x ^2} &=& -2(1- \tilde  \psi_2)\leq 0,
        \\ \label{domainIIIeq3}
  \left (\frac{\bar\xi_3}{\xi_2}\right )^2   \frac{d^2\tilde \psi_3}{d\tilde x ^2} &=& (K_{23}-\frac{\mu_3}{\bar \mu_3}) \tilde  \psi_3 \geq 0. 
   \end{eqnarray}
   The solutions in domain III read,
   \label{domainIIIsolutions}
\begin{eqnarray}
\tilde\psi_1  &=&  D_1 \,\exp{ (-\sqrt{K_{12}-1} \frac{\xi_2}{\xi_1}\tilde x) },
    \\
    \tilde \psi_2  &=&   1 - D_2 \,\exp{(-\sqrt{2} \,\tilde x)},
        \\
  \tilde \psi_3 &=& D_3 \,\exp{( -\sqrt{K_{23}-\frac{\mu_3}{\bar \mu_3}} \frac{\xi_2}{\bar\xi_3}\tilde x )}.
   \end{eqnarray}

Note that when the argument of the square root in \eqref{domainIIsolutions1} or \eqref{domainIIsolutions2} is negative, the solutions $\tilde \psi_1$ and $\tilde \psi_2$ remain valid and real. The $B$ and $C$ are then complex. Also note that $\tilde\psi_3$ features points of inflection at the domain junctions, where its second derivative changes sign (in general in a discontinuous manner). 

Due to the translational symmetry of the configurations along $\tilde x$ one may locate the origin of the coordinate system, $\tilde x=0$, at the intersection of components 1 and 2, so $\tilde \psi_1 (0) = \tilde \psi_2 (0)$. Next, since the first integral \eqref{first integral} is constant throughout, the DPA is required to preserve this property. This can be combined with the requirement that the wave functions and their first derivatives be continuous everywhere, including at the domain junctions $\tilde x^-$ and $\tilde x^+$. Since the ``kinetic" part $E_{\rm kin}$ of \eqref{first integral} is the same function in all three domains, continuity of the first derivatives of the wave functions implies continuity of $E_{\rm kin}$. The demand to keep $E_{\rm kin}+V$ constant can thus be reduced to imposing continuity of $V^{\rm(DPA)}(\tilde\psi_1(\tilde x),\tilde\psi_2(\tilde x),\tilde\psi_3(\tilde x))$ at the domain junctions $\tilde x=\tilde x^-$ and $\tilde x=\tilde x^+$. Consequently, and this is a point of special attention, the domain junctions of the DPA, at $\tilde x^-$ and $\tilde x^+$, will in general not coincide with intersection points of the wave functions of different components. An exception is a system at three-phase coexistence and at strong segregation of components 1 and 2, i.e., for $K_{12} \rightarrow \infty$. For that case, the domain junctions coincide with wave function crossings.  

\begin{figure}[htp]
    \centering
    \includegraphics[width=0.45\textwidth]{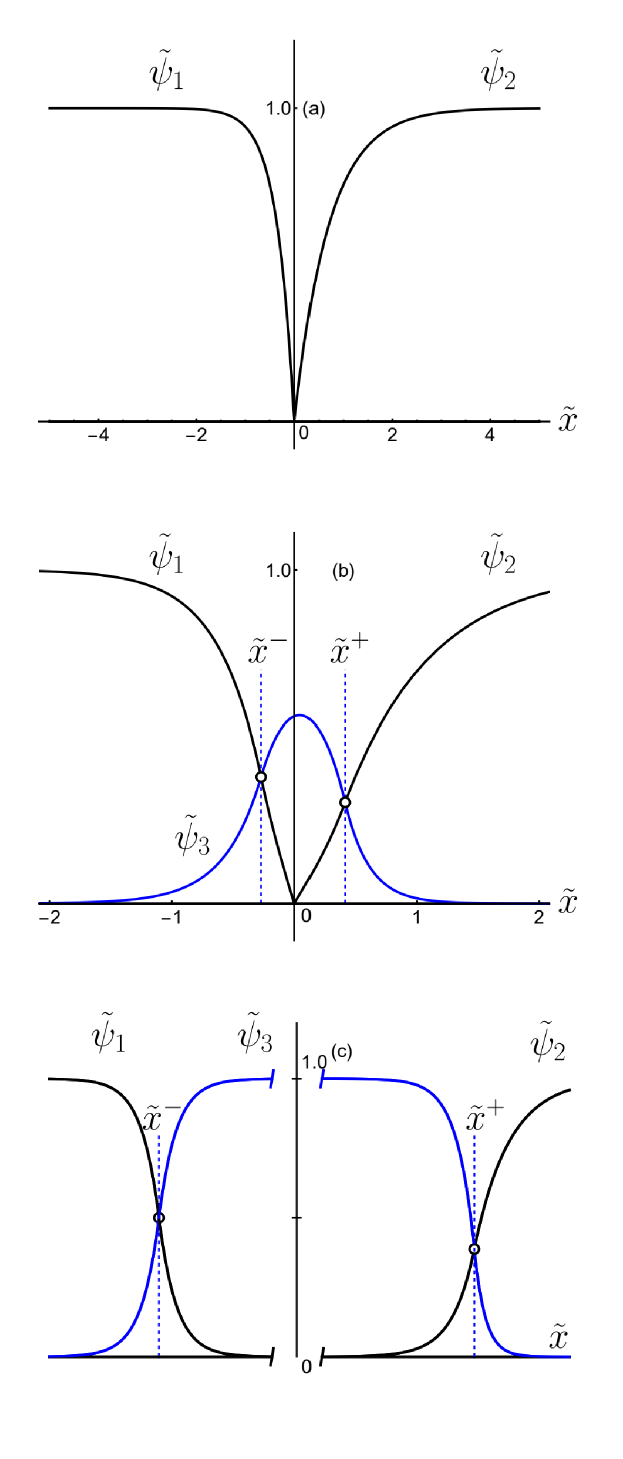}
    \caption{Interfacial wave function profiles $\tilde\psi_i$, $i=1,2,3$, according to \eqref{domainIeqsF} - \eqref{domainIIIsolutionsF}, calculated in the limit of strong segregation, $K_{12} \rightarrow \infty$, of components 1 and 2, for $\xi_2/\xi_1=2$, $\xi_3/\xi_1=1$, and at three-phase coexistence $\mu_3/\bar \mu_3 = 1$. Subfigures (a) - (c) provide quantitative examples of wave function profiles for nonwet and wet states encountered along the $x$-axis indicated in Fig.\ref{fig:trapconfig}(a), (b), and (c), respectively. (a) (Nonwet) Equilibrium 1-2 interface for $ K_{13}=5$ and $K_{23}=2K_{13}$. (b) (Nonwet) Equilibrium 1-2 interface with an adsorbed film of 3 of thickness $\tilde L \equiv \tilde x^+ - \tilde x^-$,  for $K_{13}=3.698$ and $K_{23}=2K_{13}$. In this case the wave function intersections (open circles) coincide with the domain junctions $\tilde x^- = -0.27$ and $\tilde x^+ = 0.41$.  (c)  (Wet) Equilibrium 1-2 interface wet by 3, for $K_{13}=3$ and $K_{23}=2K_{13}$. In this case $\tilde L = \infty$.}
    \label{fig:nonwetandwetinterface}
\end{figure}

In practice, to obtain the DPA wave functions, the following 15 constraints need to be satisfied. Continuity of the three wave functions, and of their first derivatives, is required at the points $\tilde x^-$ and $\tilde x^+$ (12 constraints). This permits the expression of  the 12 unknown coefficients $A_1$ to $D_3$ in terms of the remaining unknowns $\tilde x^+$ and $\tilde x^-$.  Continuity is required of $V^{(\rm DPA)}$, represented in the three domains by the functions \eqref{ternaryDPAforVdomainI}, \eqref{ternaryDPAforVdomainII}, and \eqref{ternaryDPAforVdomainIII}, respectively. That is, $V^{(\rm DPA)}_{\rm I} = V^{(\rm DPA)}_{\rm II}$ at $\tilde x^-$ and $V^{(\rm DPA)}_{\rm II} = V^{(\rm DPA)}_{\rm III}$ at $\tilde x^+$ (2 constraints). These constraints turn out to be dependent, and  determine the difference $\tilde L \equiv \tilde x^+ - \tilde x^-$, which represents a surfactant film thickness or a wetting layer thickness. Finally, fixing the origin of $\tilde x$ by requiring $\tilde\psi_1(0)=\tilde\psi_2(0)$ permits the determination of $\tilde x^+$ and $\tilde x^-$ separately. Note that in the limit of strong segregation of components 1 and 2, there are less constraints and less unknowns (see section V for details).

Let us now provide illustrations of DPA wave functions relevant to the configurations shown in Fig.~\ref{fig:trapconfig}. For simplicity, assume three-phase coexistence and consider the strong segregation limit $K_{12} \rightarrow \infty$. Then 1 and 2 are mutually impenetrable, while 1 and 3, and also 2 and 3, are mutually penetrable. Resulting nonwet and wet wave function profiles are shown in Fig.~\ref{fig:nonwetandwetinterface}.

\section{Wetting phase diagrams for weak segregation: $K_{ij} \gtrsim 1$}
In their interesting paper~\cite{JS}, Jimbo and Saito discuss the possibility that component 3 forms an equilibrium surfactant layer at the interface of components 1 and 2, thereby lowering the 1-2 interfacial tension from $\gamma_{12}$ ($\sigma_{\rm binary}(g_{12})$ in~\cite{JS}) to $\gamma_{12(3)}$ ($\sigma$ in~\cite{JS}). Wetting phase transitions are not studied in~\cite{JS}. Here, the static properties studied in~\cite{JS} are reinterpreted and complemented with the wetting phase diagram. 

\begin{figure}[htp]
    \centering
    \includegraphics[width=0.9\linewidth]{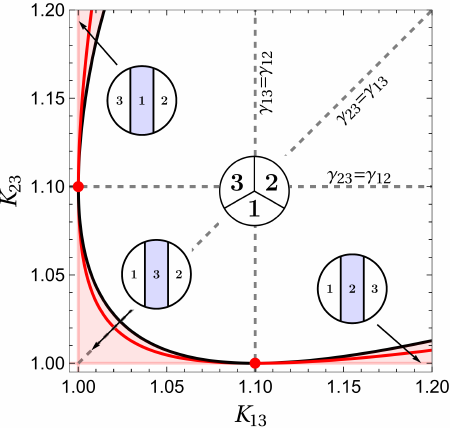}
    \caption{Wetting phase diagram in the $(K_{13},K_{23})$-plane for weak mutual segregation of all three components, $K_{ij} \gtrsim 1$. A symmetric case is considered, $\xi_1 =\xi_2=\xi_3$, and $K_{12}=1.1$ is fixed. The inner region accommodates nonwet states and the outer left and bottom regions (in pink) harbor wet states. The wetting transition is of first order. The (inner) black lines are numerically computed phase boundaries in GP theory and the (outer) red lines close to them are analytic approximations (DPA). Also shown are ``neutral" lines on which two out of three interfacial tensions are equal. The inner cartoon (akin to Fig.~\ref{fig:trapconfig}a) shows the symmetric three-phase configuration for equal interfacial tensions. The outer cartoons (akin to Fig.~\ref{fig:trapconfig}c) point to wet states in the outer region. The properties in the different quadrants of the figure are related by a permutational symmetry. In the outer region, clockwise, the wetting phase is 2, 3, or 1. The bottom left quadrant of the figure and Fig.~4c in~\cite{JS} bear similarities and differences discussed in the text.}  
    \label{fig:JSsymmetricPhD}
\end{figure}

In a first step, the same system parameters as in~\cite{JS} are taken, assuming equal atomic masses, equal intraspecies couplings $K \equiv K_{ii}$, $i=1,2,3$ ($g \equiv g_{ii}$ in~\cite{JS}), and interspecies couplings in the range for which the three components are immiscible and display weak segregation ($K_{ij} \gtrsim 1$), in particular $K_{12} = 1.1$. Three-phase coexistence is assumed, so $\mu_3 = \bar\mu_3$. The binary interfacial tensions $\gamma_{12}$, $\gamma_{13}$ and $\gamma_{23}$, as well as the three-component interfacial tension $\gamma_{12(3)}$, are computed numerically in GP theory. Analytic approximations are also provided, using the DPA. The resulting wetting phase diagram is presented in Fig.~\ref{fig:JSsymmetricPhD} and permits the assessment of the accuracy of the DPA for this system. We adopt the convention that phase boundaries computed numerically in GP theory are presented as lines in black, and (approximate) phase boundaries calculated analytically using the DPA as lines in red.

The wetting phase diagram of Fig.~\ref{fig:JSsymmetricPhD} can be best interpreted starting from the fully symmetric configuration in which all three binary (two-component) interfacial tensions are equal. This occurs when $K_{12}=K_{13}=K_{23}$. When all three components are present in an axially symmetric trap, all three dihedral angles equal $120^{\circ}$ (central cartoon). Reducing both $K_{13}$ and $K_{23}$ the angle $\hat 3$ is decreased until, when reaching the wetting phase boundary (upper black line: numerical; lower red line: DPA), $\hat 3 =0$ and $\hat 1 = \hat 2 = 180^{\circ}$.  At the wetting phase boundary, the nonwet state (without surfactant film of 3) has the same grand potential as the wet state (with a macroscopic layer of 3 intruding at the 1-2 interface). Traversing the wetting phase boundary, the equilibrium grand potential displays a jump in its first derivative as a function of $K_{13}$
and/or $K_{23}$, implying a first-order phase transition in the interfacial state. A configuration in which 3 is the wetting phase, is depicted in the lower left cartoon in Fig.~\ref{fig:JSsymmetricPhD}. The analytic approximation (DPA) to the strongly first-order wetting phase boundary reads
\begin{eqnarray}
\label{WPBstrong1stDPA}
\begin{split}
 &\frac{\sqrt{K_{12}-1} \;(\xi_1 + \xi_2)}{\sqrt{2}+ \sqrt{K_{12}-1}}  =  \\ &\frac{\sqrt{K_{13}-1} \;(\xi_1 + \xi_3)}{\sqrt{2}+ \sqrt{K_{13}-1}} +  \frac{\sqrt{K_{23}-1}\;(\xi_2 + \xi_3)}{\sqrt{2}+ \sqrt{K_{23}-1}}\,.
 \end{split}
\end{eqnarray}

The wetting phase diagram of Fig.~\ref{fig:JSsymmetricPhD} complements the results in~\cite{JS} and provides a new interpretation. The criterion $\Delta \sigma =0$, employed in~\cite{JS}, corresponds to the condition for a strongly first-order wetting transition (from no surfactant film of 3 to a macroscopic wetting layer of 3), $\gamma_{12} = \gamma_{13} + \gamma_{23}$. The lines representing the condition $\Delta \sigma =0$ in Fig.4c in \cite{JS} {\it seemingly} terminate at $(g_{13},g_{23}) = (1.07,1)$ and $(g_{13},g_{23}) = (1,1.07)$. This is due to a deficiency in the graphical layout of that figure, which hides the continuation of the lines close to the axes $g_{23}=1$ and $g_{13}=1$. In reality, just like the wetting phase boundaries in our Fig.~\ref{fig:JSsymmetricPhD}, the lines continue up till the ``neutral" points $(g_{13},g_{23}) = (1.1,1)$ and $(g_{13},g_{23})=(1,1.1)$ where, respectively, $\gamma_{12} = \gamma_{13} $ and $\gamma_{12} = \gamma_{23}$. The fact that the lines close up at the neutral points is important and signifies that there is no ``nonwetting gap".

In the light of the recent revival of the discussion around ``critical-point wetting" \cite{Cahn} and the discovery of nonwetting gaps in density-functional theory for classical fluid three-phase equilibria \cite{IK,PR}, this feature is worth discussing here. ``Critical-point wetting", reformulated for our case at hand, is the hypothesis that, for example, when $\gamma_{23}$ is decreased towards zero, while $\gamma_{13} < \gamma_{12}$, there comes a point, {\it before} $\gamma_{23}$ vanishes, from where onwards component 3 wets the 1-2 interface. Our results show that for the parameters of this phase diagram there is no nonwetting gap and the analogue of ``critical-point wetting" holds true.

\begin{figure}[htp]
    \centering
    \includegraphics[width=0.9\linewidth]{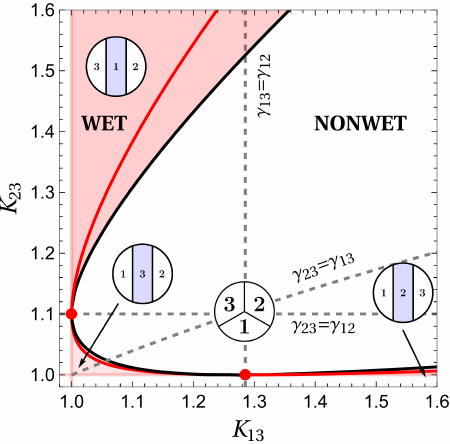}
    \caption{Wetting phase diagram in the $(K_{13},K_{23})$-plane for weak mutual segregation of all three components, $K_{ij} \gtrsim 1$. An asymmetric mixture is considered, with $\xi_2/\xi_1=2$ and $\xi_3/\xi_1=1$, and $K_{12}=1.1$ is fixed. The inner region accommodates nonwet states and the outer left and bottom regions (in pink) harbor wet states. The wetting transition is of first order. The (inner) black lines are numerically computed phase boundaries in GP theory and the (outer) red lines close to them are analytic approximations (DPA). Also shown are ``neutral" lines on which two out of three interfacial tensions are equal. The inner cartoon (akin to Fig.~\ref{fig:trapconfig}a) shows the symmetric three-phase configuration for equal interfacial tensions. The outer cartoons (akin to Fig.~\ref{fig:trapconfig}c) point to wet states in the outer region. Clockwise, the wetting phase is 2, 3, or 1.}  
    \label{fig:JSasymmetricPhD}
\end{figure}

In order to test the sensitivity of the weak-segregation wetting phase diagram to changes in the atomic constants a variant with asymmetric healing lengths is probed: $\xi_2/\xi_1 =2$ and $\xi_3/\xi_1 = 1$. The result is shown in Fig.~\ref{fig:JSasymmetricPhD}. The strongly first-order wetting transition scenario is found to be robust to this modification, and again the analogue of ``critical-point wetting" holds true.

\section{Wetting phase diagram for strong segregation of components 1 and 2}
In the strong segregation limit, $K_{12} \rightarrow \infty$, it is possible to derive an approximate wetting phase diagram analytically in great detail within the DPA.  
Since components 1 and 2 are mutually exclusive and their wave functions do not overlap, one has
\begin{eqnarray}
    \tilde \psi_1 >0, \;\; \tilde \psi_2 = 0, \;\; \mbox{for}\; \tilde x <0, \\
    \tilde \psi_2 >0, \;\; \tilde \psi_1 = 0, \;\; \mbox{for}\; \tilde x >0.
\end{eqnarray}

The DPA equations of motion, for domain I ($\tilde x < \tilde x^-$), then simplify from \eqref{domainIeq1} - \eqref{domainIeq3} to 
\begin{eqnarray}
\label{domainIeqsF}
\left (\frac{\xi_1}{\xi_2}\right )^2    \frac{d^2\tilde \psi_1}{d\tilde x ^2}  &=& -2(1- \tilde  \psi_1) \leq 0, 
    \\
       \tilde \psi_2&=& 0, 
       \\
  \left (\frac{\bar\xi_3}{\xi_2}\right )^2   \frac{d^2\tilde \psi_3}{d\tilde x ^2} &=& (K_{13}-\frac{\mu_3}{\bar \mu_3}) \tilde  \psi_3 \geq 0.
\end{eqnarray}
The solutions in domain I read,
\begin{eqnarray}
\label{domainIsolutionsF}
\tilde\psi_1  &=&  1 - A_1 \,\exp{(\sqrt{2} \frac{\xi_2}{\xi_1}\tilde x)},
    \\
    \tilde \psi_2  &=&   0,
        \\
  \tilde \psi_3 &=& A_3 \,\exp{( \sqrt{K_{13}-\frac{\mu_3}{\bar \mu_3}} \frac{\xi_2}{\bar\xi_3}\tilde x) }. 
   \end{eqnarray}

For domain II ($\tilde x^- < \tilde x < \tilde x^+ )$ the DPA equations simplify from \eqref{domainIIeq1} - \eqref{domainIIeq3} to
\begin{eqnarray}
\label{domainIIeqsF}
\left (\frac{\xi_1}{\xi_2}\right )^2    \frac{d^2\tilde \psi_1}{d\tilde x ^2}  &=& (\frac{\mu_3}{\bar \mu_3}K_{13}-1) \tilde  \psi_1, \; \mbox{for} \;\tilde x <0,
    \\
    \frac{d^2\tilde \psi_2}{d\tilde x ^2} &=& (\frac{\mu_3}{\bar \mu_3}K_{23}-1) \tilde  \psi_2,\; \mbox{for} \;\tilde x >0,
        \\
  \left (\frac{\bar\xi_3}{\xi_2}\right )^2   \frac{d^2\tilde \psi_3}
{d\tilde x ^2} &=& -2  \,\frac{\mu_3}{\bar \mu_3}( \sqrt{\frac{\mu_3}{\bar \mu_3}} - \tilde  \psi_3) \leq 0. 
   \end{eqnarray}
The solutions in domain II read, taking into account the continuity of $\tilde \psi_1$ and $\tilde \psi_2$ at $\tilde x  = 0$,
\begin{eqnarray}
\label{domainIIsolutionsF}
\tilde \psi_1  &=&   2 B_1 \,\sinh (\sqrt{\frac{\mu_3}{\bar \mu_3}K_{13}-1} \,\frac{\xi_2}{\xi_1}\tilde x ), \; \mbox{for} \; \tilde x < 0, 
    \\
    \tilde \psi_2  &=&   -2C_2 \,\sinh (\sqrt{\frac{\mu_3}{\bar \mu_3}K_{23}-1} \,\tilde x ), \; \mbox{for} \; \tilde x > 0, 
        \\
  \tilde \psi_3 &=& \sqrt{\frac{\mu_3}{\bar \mu_3}} + B_3 \,\exp{(\sqrt{2} \sqrt{\frac{\mu_3}{\bar \mu_3}} \frac{\xi_2}{\bar\xi_3}\tilde x )}\nonumber \\&+& C_3\, \exp{(-\sqrt{2} \sqrt{\frac{\mu_3}{\bar \mu_3}} \frac{\xi_2}{\bar\xi_3}\tilde x )}.
   \end{eqnarray}
   For domain III ($\tilde x > \tilde x^+$) the DPA equations simplify from \eqref{domainIIIeq1} - \eqref{domainIIIeq3} to
\begin{eqnarray}
\label{domainIIIeqsF}
\tilde \psi_1 &=& 0, 
    \\
    \frac{d^2\tilde \psi_2}{d\tilde x ^2} &=& -2(1- \tilde  \psi_2) \leq 0,
        \\ 
  \left (\frac{\bar\xi_3}{\xi_2}\right )^2   \frac{d^2\tilde \psi_3}{d\tilde x ^2} &=& (K_{23}-\frac{\mu_3}{\bar \mu_3}) \tilde  \psi_3 \geq 0, 
   \end{eqnarray}
   The solutions in domain III read,
\begin{eqnarray}
\tilde\psi_1  &=&  0,
    \\
    \tilde \psi_2  &=&   1 - D_2 \,\exp{(-\sqrt{2} \,\tilde x)}, 
        \\ \label{domainIIIsolutionsF}
  \tilde \psi_3 &=& D_3 \,\exp{( -\sqrt{K_{23}-\frac{\mu_3}{\bar \mu_3}} \frac{\xi_2}{\bar\xi_3}\tilde x )}.
   \end{eqnarray}
   
Recall that, in general, $\tilde\psi_1$ and $\tilde\psi_3$ (or $\tilde\psi_2$ and $\tilde\psi_3$) do not intersect exactly at  $\tilde x^-$ (or $\tilde x^+)$. However, at three-phase coexistence, for $\frac{\mu_3}{\bar \mu_3} =1$, and in the limit $K_{12} \rightarrow \infty$, they do. 

\begin{figure}[htp]
    \centering
    \includegraphics[width=0.9\linewidth]{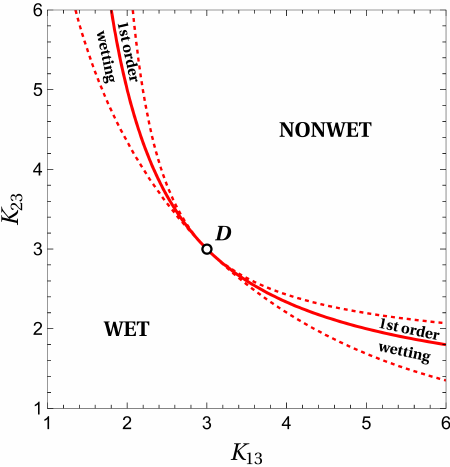}
    \caption{Wetting phase diagram in the $(K_{13},K_{23})$-plane in the strong segregation limit of condensates 1 and 2 ($K_{12} \rightarrow \infty$), for symmetric healing length ratios $\xi_2/\xi_1=1$ and $\xi_3/\xi_1=1$, calculated analytically in DPA. For strong (weak) interspecies repulsion the nonwet (wet) configuration is stable. The wetting phase transition (thick solid line; red) is strongly first-order. Auxiliary lines are drawn (dotted lines; red) that indicate the metastability limits of the wet and nonwet states. The point D indicates a degenerate first-order wetting transition.}
    \label{fig:globalPhDsym}
\end{figure}

\begin{figure}[htp]
    \centering
    \includegraphics[width=0.9\linewidth]{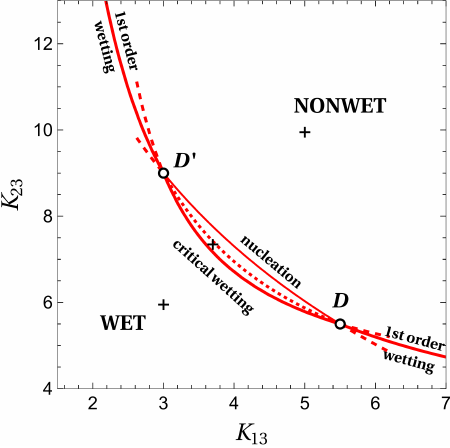}
    \caption{Wetting phase diagram in the $(K_{13},K_{23})$-plane in the strong segregation limit of condensates 1 and 2 ($K_{12} \rightarrow \infty$), for asymmetric healing length ratios $\xi_2/\xi_1=2$ and $\xi_3/\xi_1=1$, calculated analytically in DPA. For strong (weak) interspecies repulsion the nonwet (wet) configuration is stable. The wetting phase transition (thick solid line; red) is of strongly first-order for $K_{23}/K_{13} >3$ and $K_{23}/K_{13} <1$, whereas critical wetting takes place for $1 < K_{23}/K_{13} <3$. Upon lowering $K_{13}$ or $K_{23}$, critical wetting is preceded by the nucleation of a film of condensate 3 (thin solid line; red). Mathematical extensions (dashed and dotted lines; red) indicate that the wetting phase boundary displays corner singularities at the degenerate first-order wetting transitions at D and D'. The three points marked ``$+$" locate, for descending $K_{13}$ and fixed $K_{23}/K_{13}=2$, the state points associated with the interface configurations shown in Fig.~\ref{fig:nonwetandwetinterface}a-c, calculated analytically within DPA.}
    \label{fig:globalPhDasym}
\end{figure}

As in the previous section, the interspecies scattering lengths are varied so that the control parameters are $K_{13}$ and $K_{23}$, and the healing length ratios are fixed first symmetrically, $\xi_2/\xi_1 = \xi_3/\xi_1 = 1$ and, next, asymmetrically, $\xi_2/\xi_1 =2$ and $\xi_3/\xi_1 = 1$. The wetting phase transitions and critical phenomena so uncovered belong to three distinct classes: strongly first-order wetting with an energy barrier, a borderline case of degenerate first-order wetting (without energy barrier) and critical wetting. The wetting phase diagrams at three-phase coexistence are shown in Fig.~\ref{fig:globalPhDsym} and Fig.~\ref{fig:globalPhDasym}, respectively, for the symmetric and the asymmetric mixture.

In general, a first-order phase transition is accompanied by an energy barrier which separates two energy minima with equal values of the energy. In the context of wetting transitions, one often defines an ``interface potential" $V(\tilde L)$, which is a generalization of the grand potential to non-equilibrium states with arbitrary film thickness $\tilde L$. The equilibrium grand potential $\Omega$ is then found as the minimum of $V(\tilde L)$. The interface potential is calculated using $\tilde L$ as a constraint and minimizing a generalized grand potential in which the disjoining pressure acts as a Lagrange multiplier. This scheme has been outlined in great detail in \cite{VSIc} for a two-component BEC at an optical wall. 

Presently there is no need to calculate $V(\tilde L)$. The character of the wetting transitions is interpreted and elucidated using the concepts and insights gained from the interface potential calculations for wetting developed in \cite{VSNI}.
At first-order wetting $V(0)$ and $V(\infty)$ are equal minima of $V(\tilde L)$ with an energy barrier in between. The height of the barrier can be found by calculating the energy of the unstable state that also solves the GP equations (here in DPA). This unstable state corresponds conceptually to the maximum of $V(\tilde L)$. 

For the symmetric case (Fig.~\ref{fig:globalPhDsym}) the wetting transition is strongly of first order. The equilibrium wetting layer thickness $\tilde L$ jumps from zero to a macroscopic (``infinite") value. The point $D$ denotes a degenerate first-order transition. At $D$, states with a surfactant layer of 3 are all equilibrium states, regardless of their thickness $\tilde L$, as they all have the same value of the grand potential. This degeneracy is similar to that found for the strongly first-order wetting transition for two condensates at a hard wall \cite{IVS}, and is analysed in detail with the aid of the interface potential calculations in \cite{VSNI}; at $D$, $V(\tilde L)$ is independent of $\tilde L$. Away from $D$ the strongly first-order wetting transition features an energy barrier, between the energy minima at $\tilde L =0$ and $\tilde L = \infty$. The dotted lines in the phase diagram denote the metastability limit of the metastable states (local minima of the energy). We return to these in more detail in the asymmetric case.

\begin{figure*}[htp]
    \centering
    \begin{subfigure}{0.485\linewidth}
        \includegraphics[width=0.80\linewidth]{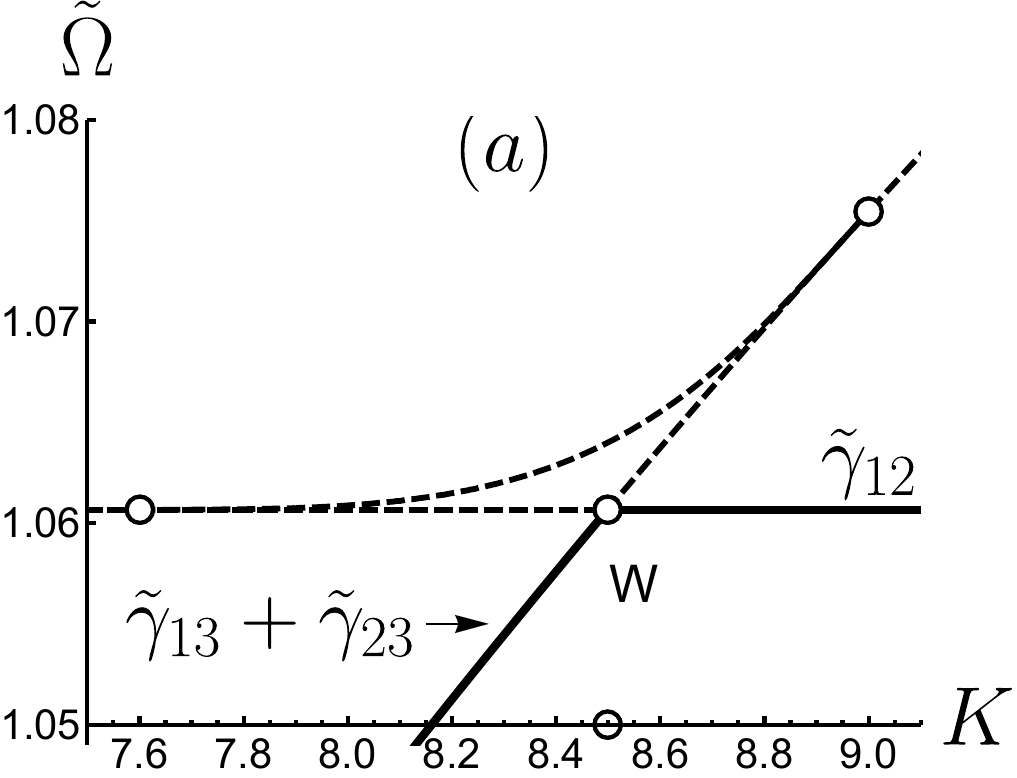}
    \end{subfigure}
    \begin{subfigure}{0.485\linewidth}
        \includegraphics[width=0.80\linewidth]{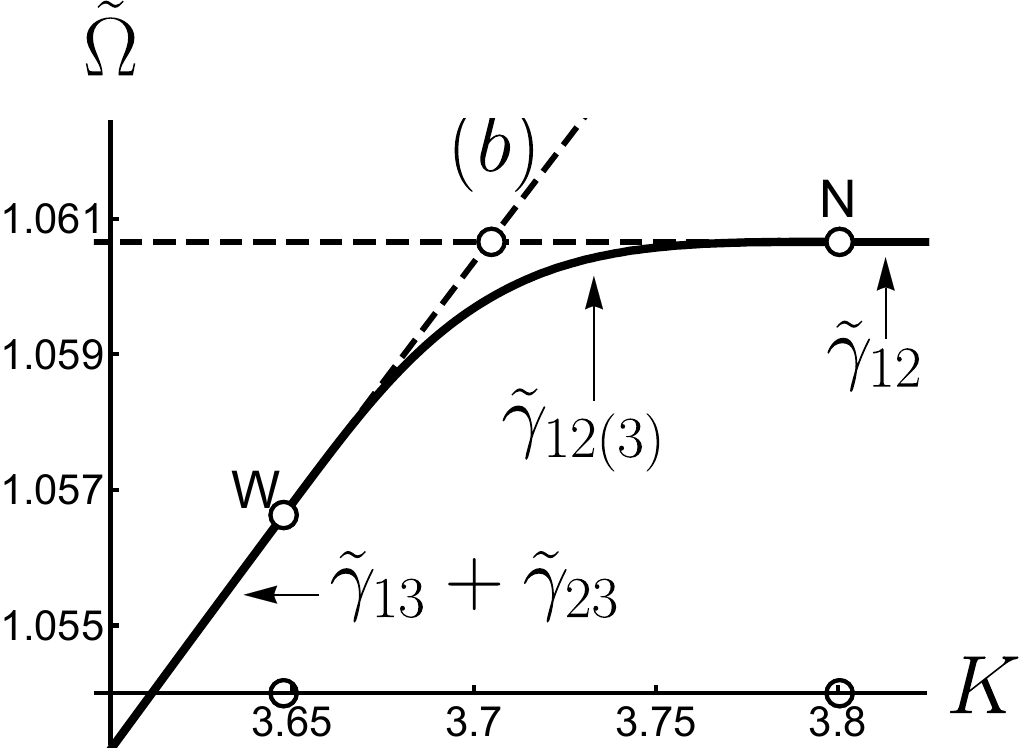}
    \end{subfigure}
    \caption{Reduced surface excess grand potential $\tilde \Omega \equiv \Omega/ 4P\xi_2$ versus $ K\equiv K_{13}$ for equilibrium and non-equilibrium states. Two straight paths are followed in the phase diagram of Fig.\ref{fig:globalPhDasym}. (a) Strongly first-order wetting transition with an energy barrier, for the path $K_{23}/ K_{13} = 0.5$. Shown are the branches of $\tilde \Omega$ corresponding to the minimum of $\tilde \Omega$ for a given value of $ K$ (equilibrium state; thick solid lines), intermediate values of $\tilde \Omega$ (metastable states; dashed straight lines) and the maximum of $\tilde \Omega$ (unstable state; dashed curve). The left-most (right-most) open dot is the metastability limit of the nonwet (wet) state, and the dot at W indicates the wetting transition. (b) Critical wetting transition, for the path $K_{23}/ K_{13} = 2$. The branches of $\tilde \Omega$ corresponding to the minimum of $\tilde \Omega$  for a given value of $ K$ give the equilibrium states (thick solid lines). A thin film of 3 is nucleated at N and the equilibrium layer thickness $\tilde L$ increases continuously with decreasing $K$. Approaching the critical wetting transition W, $\tilde L$ diverges in a continuous manner.}
    \label{fig:reducedOmega}
\end{figure*}

For the asymmetric case (Fig.~\ref{fig:globalPhDasym}), in two outer sectors, $K_{23} < K_{13}$ and $K_{23} > 3 K_{13}$, the wetting transition is strongly of first order. The equilibrium wetting layer thickness $\tilde L$ jumps from zero to a macroscopic (``infinite") value. The slope of the equilibrium $\Omega$ versus $K_{13}$ is discontinuous at the wetting transition, where the equilibrium $\gamma_{12(3)}$ crosses over from $\gamma_{12}$ to $\gamma_{13}+\gamma_{23}$. This is illustrated in Fig.~\ref{fig:reducedOmega}a for a path at constant ratio $K_{23}/ K_{13}$.

In contrast, in the inner sector of the phase diagram (Fig.~\ref{fig:globalPhDasym}), for $K_{13} < K_{23} < 3K_{13}$,
 {\it en route} to the wetting transition, an equilibrium  wetting layer of condensate 3 of finite thickness $\tilde L$ develops. It originates at a nucleation transition, a quantum phenomenon in which an infinitesimal $\tilde \psi_3$ is created. Decreasing the interspecies atomic repulsive forces, the equilibrium value of $\tilde L$ increases to a macroscopic value and, theoretically, diverges at the wetting point. This divergence is logarithmic as expected for systems with exponentially decaying surface forces \cite{Dietrich,Parry}. Plotting the surface excess grand potential as a function of $K_{13}$ at constant $K_{23}/ K_{13}$ leads to Fig.~\ref{fig:reducedOmega}b. The slope of the equilibrium $\Omega$ is continuous at W, hence the name continuous wetting or ``critical" wetting. 

At the special points D and D' in the phase diagram nucleation and wetting coincide. This renders the wetting transition degenerate: the grand potential is independent of the wetting layer thickness (cf. the discussion of the point $D$ for the symmetric case). 

For $K_{12}\rightarrow\infty$, the DPA permits one to find simple analytic approximations for the phase boundaries. The nucleation line, found by studying the onset of stability of an infinitesimal film of 3 at the 1-2 interface, satisfies 
\begin{equation}
\label{nucleationline}
    \xi_1+\xi_2 =\left (\sqrt{K_{13}-1}+\sqrt{K_{23}-1}\right)\frac{\xi_3}{\sqrt{2}}\,.
\end{equation}
The strongly first-order wetting phase boundary, obtained by requiring $\gamma_{12} = \gamma_{13} + \gamma_{23}$ (no surfactant), reads
\begin{equation}
\label{WPBhard12interface}
\begin{split}
 \xi_1 + \xi_2  =  \frac{\sqrt{K_{13}-1} \;(\xi_1 + \xi_3)}{\sqrt{2}+ \sqrt{K_{13}-1}}+  \frac{\sqrt{K_{23}-1}\;(\xi_2 + \xi_3)}{\sqrt{2}+ \sqrt{K_{23}-1}}\,,
 \end{split}
\end{equation}
which is simply the strong segregation limit of \eqref{WPBstrong1stDPA}.

The critical wetting phase boundary, derived by asymptotic analysis, for $\tilde L \rightarrow \infty$, of $\gamma_{12(3)}$ and by imposing the equality in \eqref{Antonov}, obeys
\begin{equation}
\label{truewet}
\begin{split}
   \frac{\xi_1}{\sqrt{K_{13}-1}} + \frac{\xi_2}{\sqrt{K_{23}-1}} = \sqrt{2}\;\xi_3.
\end{split}
\end{equation}
To our knowledge, this is the first occasion on which an analytic result is obtained for a {\it critical} wetting transition in mixtures of BEC.

\section{Wetting phase diagram for intermediate segregation}
In the weak-segregation regime, in which all three condensates pairwise interpenetrate significantly, strongly first-order wetting is found from numerical solution of the GP equations. In the strong segregation limit, on the other hand, the DPA suggests that besides strongly first-order wetting, a range of parameters exists in which critical wetting occurs. It is therefore necessary to explore, in the intermediate regime, which type of wetting transition can generically be expected, when solving the full GP equations numerically.

\begin{figure}
    \centering
    \includegraphics[width=0.9\linewidth]{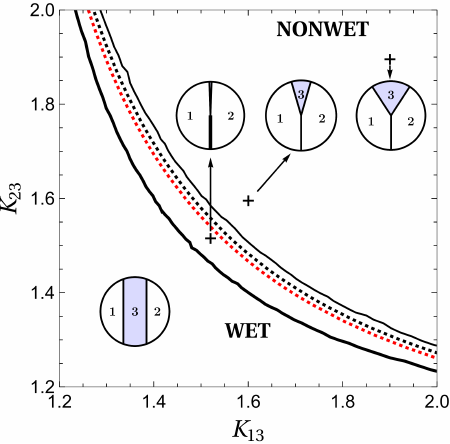}
    \caption{Wetting phase diagram in the $(K_{13},K_{23})$-plane for symmetric healing length ratios $\xi_2/\xi_1=1$ and $\xi_3/\xi_1=1$, and for intermediate segregation between condensates 1 and 2 ($K_{12} = 10$), computed numerically by solving the GP equations \eqref{coupledGP}. For strong (weak) interspecies repulsion the equilibrium state is the nonwet (wet) configuration. The wetting phase transition is critical and its phase boundary, which satisfies $\gamma_{12(3)} = \gamma_{13} + \gamma_{23}$, is shown (thick solid line; black). Upon lowering $K_{13}$ or $K_{23}$ from a nonwet state, critical wetting by condensate 3 is preceded by the nucleation of a film of condensate 3 (thin solid line; black). The auxiliary line (dotted line; black) is the locus of the equality $\gamma_{12} = \gamma_{13} + \gamma_{23}$, which here does not describe the wetting transition because it neglects the equilibrium film of 3 at the 1-2 interface. The DPA for this auxiliary line, which obeys \eqref{WPBstrong1stDPA}, is also shown (dotted line; red). The three points marked ``$+$" correspond to nonwet states along the line $K_{13}=K_{23}$. The cartoons display precise dihedral angles near the three-phase contact line, for these three state points. The angles, obtained from the numerically computed interfacial tensions using \eqref{coshat3} (and its cyclic permutations) are $\hat 1 = \hat 2 = 147^{\circ}$ and $\hat 3 = 66^{\circ}$ for $K_{13} = 1.9$, $\hat 1 = \hat 2 = 164^{\circ}$ and $\hat 3 = 32^{\circ}$ for $K_{13} = 1.6$, and $\hat 1 = \hat 2 = 178^{\circ}$ and $\hat 3 = 4^{\circ}$ for $K_{13} = 1.52$. In this last case a thin film of 3 is present at the 1-2 interface (as depicted in the cartoon).}
        \label{fig:symIntermediate}
\end{figure}

\begin{figure}
    \centering
    \includegraphics[width=0.9\linewidth]{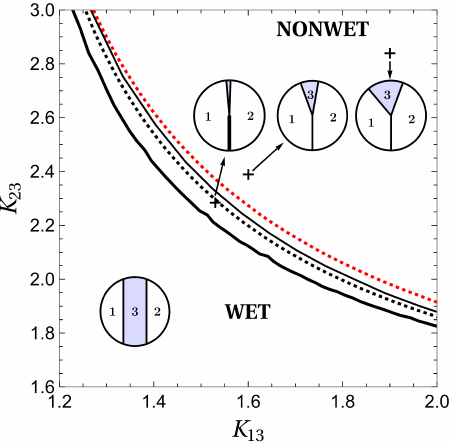}
    \caption{ Wetting phase diagram in the $(K_{13},K_{23})$-plane for asymmetric healing length ratios $\xi_2/\xi_1=2$ and $\xi_3/\xi_1=1$, and for intermediate segregation between condensates 1 and 2 ($K_{12} = 10$), computed numerically by solving the GP equations \eqref{coupledGP}. The description of the lines in the figure is identical to that in the caption for the symmetric healing length ratios case Fig.~\ref{fig:symIntermediate}. The three points marked ``$+$" correspond to nonwet states along the line $K_{23} = (3/2) K_{13}$. The cartoons display precise dihedral angles near the three-phase contact line, for these three state points. The angles, obtained from the numerically computed interfacial tensions using \eqref{coshat3} (and its cyclic permutations) are $\hat 1 = 140^{\circ}$, $\hat 2 = 160^{\circ}$ and $\hat 3 = 60^{\circ}$ for $K_{13} = 1.9$, $\hat 1 = 160^{\circ}$, $\hat 2 = 170^{\circ}$ and $\hat 3 = 30^{\circ}$ for $K_{13} = 1.6$, and $\hat 1 = 175^{\circ}$, $\hat 2 = 178^{\circ}$ and $\hat 3 = 7^{\circ}$ for $K_{13} = 1.53$. In this last case a thin film of 3 is present at the 1-2 interface (as depicted in the cartoon).}
    \label{fig:asymIntermediate}
\end{figure}

Wetting phase diagrams were obtained numerically for $K_{12} = 10^n$, with $n=1,2, ..., 5$. In all cases {\it critical wetting} was found. To illustrate these results, the wetting phase diagram is presented for $K_{12} = 10$, for the symmetric and for an asymmetric choice of healing length ratios, respectively, in Fig.~\ref{fig:symIntermediate} and Fig.~\ref{fig:asymIntermediate}. In these phase diagrams the wetting phase boundary, together with auxiliary lines, are shown. The main auxiliary line is the nucleation transition, from $\tilde \psi_3 =0$ to an infinitesimal $\tilde \psi_3$. Furthermore, the auxiliary line, here not a phase transition, on which $\gamma_{12} = \gamma_{13}+ \gamma_{23}$ is satisfied, is shown, together with its DPA counterpart. Again, this gives us a means to assess the accuracy of the DPA in a typical circumstance.

A remarkable property of the fluid three-phase equilibria at hand is that the dihedral angles in a two-dimensional cross section of a nonwet state can be obtained from a force balance among interfacial tensions calculated far away from the contact line assuming only a one-dimensional inhomogeneity (in mean field, ignoring fluctuations). In this vein, note that the dihedral angles are determined by the interfacial tensions through force balance equations given in detail in \cite{RW}. In particular, one has
\begin{equation}\label{coshat3}
\cos\hat 3 = \frac{1}{2}\left ( \frac{\sigma_{12(3)}}{\sigma_{13}} \frac{\sigma_{12(3)}}{\sigma_{23}} - \frac{\sigma_{13}}{\sigma_{23}} - \frac{\sigma_{23}}{\sigma_{13}}\right ),
\end{equation}
and suitable cyclic permutations determine $\hat 1$ and $\hat 2$.
To illustrate the calculation of dihedral angles from interfacial tensions, the angles are depicted (in the cartoons), and their numerical values reported (in the captions), for various nonwet states at three-phase coexistence, in Fig.~\ref{fig:symIntermediate} and Fig.~\ref{fig:asymIntermediate}.


Additional evidence for the occurrence of critical wetting is provided by examining the states off of three-phase coexistence. In contrast with a first-order wetting transition, for a critical wetting transition there is no prewetting line attached to it in the phase diagram. However, a nucleation line (for condensate 3) is expected, which meets bulk three-phase coexistence at a nucleation point $N$ distinct from the critical wetting point $W$. In the next section these phenomena are investigated in the weak, strong and intermediate segregation regimes.

\section{Prewetting phenomena off of three-phase coexistence}
\begin{figure}[htp]
    \centering
    \includegraphics[width=0.4\linewidth]{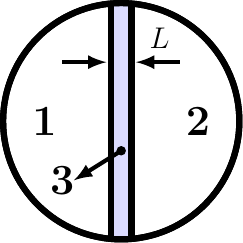}
    \caption{Three-component BEC configuration off of three-phase coexistence, in a prewetting state. Condensates 1 and 2 are at two-phase coexistence ($P_1 = P_2 = P$) while condensate 3 is not stable in bulk ($P_3 < P$) but is present as a prewetting film at the 1-2 interface. When $P_3$ is increased to $P$, the film thickness $L$ diverges and complete wetting is achieved (cf. Fig.1c).}
    \label{fig:prewettingtrapconfig}
\end{figure}

In the grand canonical ensemble, and considering states in which condensates 1 and 2 are at two-phase coexistence but condensate 3 is not stable as a bulk phase, i.e., $\mu_3 < \bar \mu_3$, there are two possibilities. The first (i) is that component 3 is not present at the 1-2 interface, in which case $\tilde \psi_3 (\tilde x)=0$ for all $\tilde x$. The second (ii) is that component 3 forms an equilibrium surfactant layer of nanoscopic thickness $\tilde L$ at the 1-2 interface, resulting in a lower 1-2 interfacial tension \cite {JS}.
This case is illustrated in Fig.~\ref{fig:prewettingtrapconfig} and the configuration is commonly referred to as a ``prewetting state".

The transition from (i) to (ii) is a ``nucleation" transition, in which an infinitesimal $\tilde \psi_3$ appears (for a continuous nucleation transition) or a jump from zero to a finite $\tilde \psi_3$ occurs (for a first-order nucleation transition).
Nucleation transitions typically occur in systems described by quantum mechanical wave functions (BEC mixtures \cite{IVS,VSIc},  superconductors \cite{SJdG,IvL2}), and not in classical fluids at temperature $T >0$. The density or concentration of a classical component in a mixture cannot be strictly zero in some region of space. At any location, each component is present, even if only in very small amounts, due to the entropy gain of mixing (for $T>0$).

Nucleation transitions bear similarities to prewetting transitions, but there are differences that must be discussed, in our context. At nucleation a nonzero $\tilde \psi_3$ is generated, while at a prewetting transition a layer of 3 is formed (either from zero or from a thin film), the thickness $\tilde L$ of which diverges ($\tilde L \rightarrow \infty$) when three-phase coexistence ($\mu_3 \rightarrow \bar\mu_3$) is reached, resulting in a wet state. Standardly, the prewetting transition is the extension off of coexistence of a first-order wetting transition at coexistence. In that strict sense there is no prewetting transition when the wetting transition is continuous (or ``critical"), but there may well be a nucleation transition. The distinction, when it must be made, becomes clear when examining the wetting and prewetting phase diagrams.

The prewetting phase diagram also gives the location of the relevant bulk phase transitions. These are determined by examining the minimum of the grand potential for spatially homogeneous bulk states. Recall that the pure components 1 and 2 are at two-phase coexistence ($P_1 = P_2 = P$). Without loss of generality, suppose $K_{13} \leq K_{23}$. Consider first $K_{13} < K_{23}$. Then, in the plane of $K_{13}$ and $\mu_3/\bar\mu_3$, keeping $K_{12} (>1)$ and $K_{23} (>1)$ fixed, bulk three-phase coexistence occurs for $\mu_3/\bar\mu_3 =1$ and $K_{13} > 1$. Following the analysis given in \cite{VSIc} one finds that in equilibrium, upon lowering $K_{13}$ at fixed $\mu_3/\bar\mu_3 < 1$, a (critical) bulk phase transition from two coexistent phases with pure components 1 and 2 (without 3) towards a single mixed phase of 1 and 3 (without 2) takes place at 
\begin{equation}
\label{12toMixed13}
K_{13} = \frac{\mu_3}{\bar\mu_3}, \; \mbox{for}\; \mu_3 <\bar\mu_3.
\end{equation}
Conversely, in equilibrium, upon lowering $K_{13}$ at fixed $\mu_3/\bar\mu_3 > 1$, a (critical) bulk phase transition from a phase with pure component 3  towards a single mixed phase of 1 and 3 (without 2) takes place at
\begin{equation}
\label{3toMixed13}
K_{13} = \frac{\bar \mu_3}{\mu_3}, \; \mbox{for}\; \mu_3 > \bar\mu_3.
\end{equation}
For $\mu_3 = \bar\mu_3$, a first-order transition takes place from three coexistent pure phases 1, 2 and 3 to two coexistent phases, being pure 2 and a mixed phase of 1 and 3, at $K_{13}=1$ (as already found in \cite{Roberts}). Consider next $K_{13} = K_{23}$. Then the preceding reasoning applies, with the modification that the bulk transition to the mixed phase is a transition from bulk three-phase equilibrium of three pure phases (1, 2 and 3) to bulk two-phase equilibrium of a mixed phase of 1 and 3 coexisting with a mixed phase of 2 and 3. 

\begin{figure}[htp]
    \centering
    \includegraphics[width=0.95\linewidth]{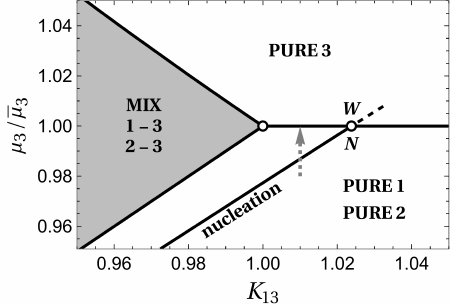}
    \caption{Prewetting phase diagram in chemical potential $\mu_3/\bar \mu_3$ versus coupling $K_{13} (= K_{23})$, for weak segregation at system parameters $K_{12} = 1.1$, $\xi_1=\xi_2=\xi_3$. The associated wetting phase diagram is Fig.~\ref{fig:JSsymmetricPhD}. 
    Bulk two-phase coexistence of pure component 1 and pure component 2 is the equilibrium state for $\mu_3/\bar \mu_3 < 1$ and $K_{13} > \mu_3/\bar \mu_3$, while a single phase of pure 3 is stable for $\mu_3/\bar \mu_3 > 1$ and $K_{13} > \bar\mu_3/ \mu_3$. In the remainder (area in grey) bulk two-phase coexistence occurs of a mixed phase of components 1 and 3 and a mixed phase of 2 and 3. Bulk three-phase coexistence takes place for $\mu_3/\bar \mu_3 = 1$ and $K_{13} >1$. At three-phase coexistence the degenerate first-order wetting transition $W$ coincides with the nucleation point $N$. The nucleation line (solid line, computed numerically in GP theory) meets bulk coexistence at an angle. The mathematical extension is shown (dashed line). The vertical path (dotted arrow) indicates the nucleation of a prewet state and the approach to a wet state.}
    \label{fig:PrewetJSsymmetric}
\end{figure}

For weak segregation ($K_{ij} \gtrsim 1$), with $K_{12} = 1.1$ fixed and a symmetric mixture ($\xi_1=\xi_2=\xi_3$), recall the wetting phase diagram of Fig.~\ref{fig:JSsymmetricPhD}. The prewetting phase diagram, computed in GP theory, associated with a strongly first-order wetting transition at $W$, located at $\mu_3/\bar\mu_3 =1, K_{13} = K_{23} \approx 1.0238$, is presented in Fig.~\ref{fig:PrewetJSsymmetric}. In this diagram $K_{13}=K_{23}$ is assumed. Note that $W$ coincides with the nucleation point $N$, which is the intersection of the numerically computed nucleation line with the bulk three-phase coexistence line $\mu_3 = \bar \mu_3$. All along this line the nucleation transition is continuous, and can be interpreted as the prewetting transition. The fact that a first-order wetting transition can be accompanied by a line of continuous prewetting transitions that does {\it not} meet the bulk coexistence line tangentially, is indicative of degenerate first-order wetting at $W$. The thermodynamic consistency of this extraordinary phenomenon has been explained in \cite{IVS}.

A useful order parameter for wetting, and more generally accessible than the layer thickness $\tilde L$, is the adsorption of component 3, defined as
\begin{equation}
    \Gamma_3 = \int_{-\infty}^{\infty} d\tilde z \; \tilde \psi_3(\tilde z)^2.
\end{equation}
For large adsorption, close to wetting, $\Gamma_3 \propto \tilde L$, and for small adsorption, close to nucleation, $\Gamma_3$ is well defined, while $\tilde L$ is not.

\begin{figure}[htp]
    \centering
    \includegraphics[width=0.95\linewidth]{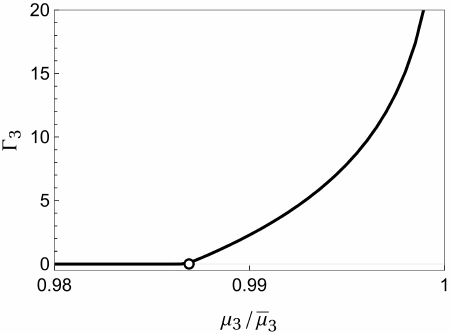}
    \caption{Adsorption $\Gamma_3$ of component 3 versus chemical potential $\mu_3/\bar\mu_3$, along the vertical path indicated in Fig.~\ref{fig:PrewetJSsymmetric}. The system parameters are $K_{12} = 1.1$, $K_{13}=K_{23} = 1.01$ and $\xi_1=\xi_2=\xi_3$. The nucleation transition (white dot) is indicated and the prewet state approaches a wet state for $\mu_3/\bar\mu_3 \uparrow 1$.}
    \label{fig:Gamma3JS}
\end{figure}

Fig.~\ref{fig:Gamma3JS} displays how $\Gamma_3$ varies along the path at constant $K_{13} = K_{23}= 1.01$ indicated by the dashed line in Fig.~\ref{fig:PrewetJSsymmetric}. Along that path, in the direction of the arrow, one encounters the nucleation transition at $\mu_3/\bar \mu_3 \approx 0.9869$ and upon approach of three-phase coexistence $\Gamma_3$ diverges as the complete wetting state is reached. This divergence is logarithmic, $\Gamma_3 \propto \ln (1/(1-\mu_3/\bar\mu_3))$, as has already been established in the context of the GP theory for two components at a wall \cite{VSIc}. Fig.~\ref{fig:Gamma3JS} is quantitatively similar to Fig.~2b in~\cite{JS}, and complements that figure, since the nucleation transition was not studied in~\cite{JS}.

In the strong segregation limit of components 1 and 2 ($K_{12} \rightarrow\infty$) an analytic approximation to the nucleation transition is derived within the DPA, under the following assumptions: i) the magnitude of $\tilde \psi_3$ is infinitesimal, and ii) at nucleation of component 3 the domain junctions $\tilde x^-$ and $\tilde x^+$ coincide at $\tilde x = 0$, implying $L \equiv \tilde x^+ - \tilde x^- \downarrow 0$. This leads, for $\mu_3 \leq \bar \mu_3$, to the following DPA for the nucleation transition,  
\begin{eqnarray}
\label{nucleationline}
    \sqrt{2}\left(\frac{\xi_1 + \xi_2}{\bar\xi_3}\right)\left(\frac{\mu_3}{\bar\mu_3}\right)^{3/2}=\sqrt{K_{13}-\frac{\mu_3}{\bar\mu_3}}+\sqrt{K_{23}-\frac{\mu_3}{\bar\mu_3}}. \nonumber \\
\end{eqnarray}

\begin{figure}[htp]
    \centering
    \includegraphics[width=0.95\linewidth]{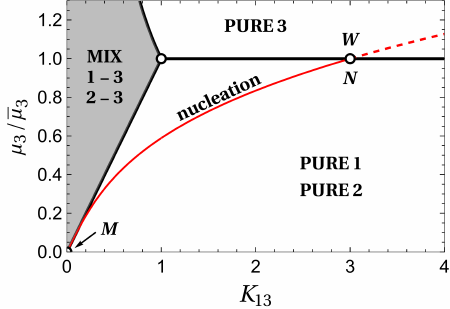}
    \caption{Prewetting phase diagram in chemical potential $\mu_3/\bar \mu_3$ versus coupling $K_{13} (= K_{23})$, for strong segregation at system parameters $K_{12} \rightarrow \infty$, $\xi_1=\xi_2=\xi_3$. The associated wetting phase diagram is Fig.~\ref{fig:globalPhDsym}. The bulk phases and transitions among them are identical to those in Fig.~\ref{fig:PrewetJSsymmetric}. At three-phase coexistence the degenerate first-order wetting transition $W$ coincides with the nucleation point $N$. The nucleation line (solid line, red) and its mathematical extension (dashed line, red) represent the DPA~\eqref{nucleationline}. The line meets the mixed phase at $M$, in the origin of the diagram.}
    \label{fig:prewettingstrongsym}
\end{figure}

\begin{figure}[htp]
    \centering
    \includegraphics[width=0.95\linewidth]{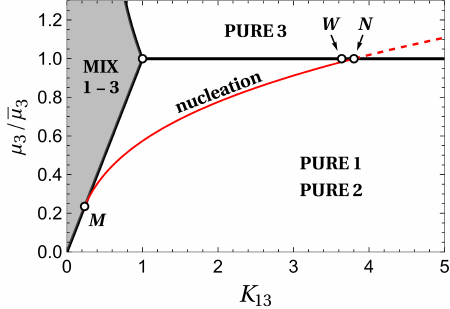}
    \caption{Prewetting phase diagram in chemical potential $\mu_3/\bar \mu_3$ versus coupling $K_{13}$, with $K_{13} = K_{23}/2$, for strong segregation at system parameters $K_{12} \rightarrow \infty$, $\xi_2/\xi_1 = 2$ and $\xi_3/\xi_1=1$. The associated wetting phase diagram is Fig.~\ref{fig:globalPhDasym}. The bulk phases and transitions among them are the same as those in Fig.~\ref{fig:PrewetJSsymmetric}, except that the mixed phase (of components 1 and 3) is now a single phase. At three-phase coexistence $W$ marks the critical wetting transition. It is separated from the nucleation line, which ends at $N$. The nucleation line (solid line, red) and its mathematical extension (dashed line, red) represent the DPA~\eqref{nucleationline}. The nucleation line meets the mixed phase at point $M$.}
    \label{fig:prewettingstrongasym}
\end{figure}

In the strong segregation limit the wetting phase diagrams calculated in DPA are Fig.~\ref{fig:globalPhDsym} and Fig.~\ref{fig:globalPhDasym}. The respective DPA prewetting phase diagrams are shown in Fig.~\ref{fig:prewettingstrongsym} for the symmetric mixture and in Fig.~\ref{fig:prewettingstrongasym} for the asymmetric one. Note that the nucleation line, given in~\eqref{nucleationline}, meets the bulk coexistence line at an angle, at the degenerate first-order wetting transition in Fig.~\ref{fig:prewettingstrongsym}, while the nucleation line passes underneath the critical wetting transition $W$ and meets the bulk coexistence line at an angle at the nucleation point $N$ in Fig.~\ref{fig:prewettingstrongasym}.


\begin{figure}[htp]
    \centering
    \includegraphics[width=0.95\linewidth]{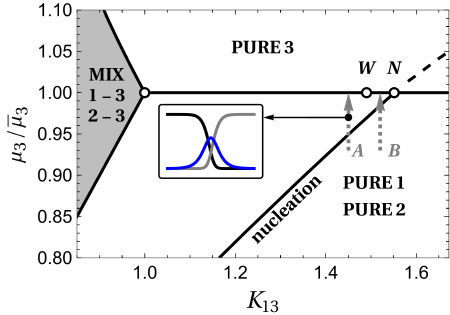}
    \caption{Prewetting phase diagram in chemical potential $\mu_3/\bar \mu_3$ versus coupling $K_{13} (= K_{23})$, for intermediate segregation at system parameters $K_{12} = 10$, $\xi_1=\xi_2=\xi_3$. The associated wetting phase diagram is Fig.~\ref{fig:symIntermediate}. 
    The bulk phases and transitions among them are identical to those in Fig.~\ref{fig:PrewetJSsymmetric}. At three-phase coexistence the critical wetting transition, at $W$, is separated from the nucleation point $N$. The nucleation line (solid line, computed numerically in GP theory) meets bulk coexistence at an angle. The mathematical extension is shown (dashed line). The vertical paths (dotted line) indicate the nucleation of a prewet state and the approach to a wet state, for path $A$, and the nucleation of a surfactant film and the approach to a nonwet state, for path $B$. For a selected state on path $A$ (black dot), the wave function profiles are displayed in the inset (for detail, see Fig.~\ref{fig:Prewetprofiles}).}
    \label{fig:PrewetInterm}
\end{figure}

\begin{figure}[htp]
    \centering
    \includegraphics[width=0.95\linewidth]{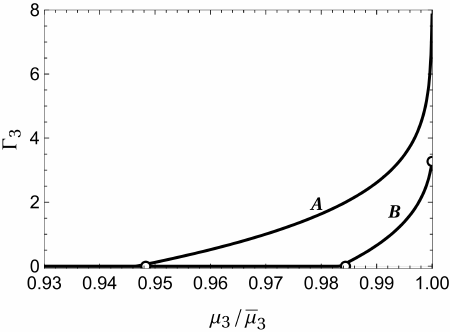}
    \caption{Adsorption $\Gamma_3$ of component 3 versus chemical potential $\mu_3/\bar\mu_3$, along the vertical paths $A$ and $B$ indicated in Fig.~\ref{fig:PrewetInterm}. The system parameters are $K_{12} = 10$, $K_{13}= K_{23}=1.45$ (path $A$), $K_{13}= K_{23}=1.52$ (path $B$) and $\xi_1=\xi_2=\xi_3$. The nucleation transitions (white dots at $\Gamma_3 = 0$) are indicated and the prewet state (path $A$) approaches a wet state with $\Gamma_3 \rightarrow\infty$ for $\mu_3/\bar\mu_3 \uparrow 1$. In contrast, path $B$ does not lead to a wet state. At three-phase coexistence the nonwet state for path $B$ features the finite adsorption $\Gamma_3 = 3.27$ (white dot at $\mu_3/\bar \mu_3=1$).}
    \label{fig:Gamma3Interm}
\end{figure}

In the intermediate segregation regime, with $K_{12}= 10$, a numerically computed prewetting phase diagram associated with the symmetric mixture wetting phase diagram of Fig.~\ref{fig:symIntermediate}, is presented in Fig.~\ref{fig:PrewetInterm}. The nucleation transition is a continuous phase transition from $\Gamma_3=0$ to an infinitesimal $\Gamma_3$. The nucleation line passes underneath the critical wetting transition $W$ and meets the bulk coexistence line at an angle at the nucleation point $N$. The adsorption $\Gamma_3$ is computed along the paths $A$, at $K_{13}=K_{23}=1.45$, and $B$, at $K_{13}=K_{23}=1.52$, approaching three-phase coexistence. The results are shown in Fig.~\ref{fig:Gamma3Interm}. 

\begin{figure}[htp]
    \centering
    \includegraphics[width=0.95\linewidth]{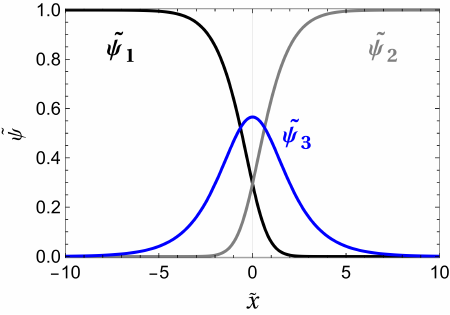}
    \caption{Interfacial wave function profiles $\tilde\psi_i$, $i=1,2,3$, computed numerically in GP theory, for a prewetting state off of three-phase coexistence corresponding to the inset in Fig.~\ref{fig:PrewetInterm}. The system parameters are $\mu_3/\bar \mu_3 =0.97$, $K_{12}=10$, $K_{13}=K_{23} = 1.45$ and $\xi_1=\xi_2=\xi_3$. For this state, $\Gamma_3 \approx 1.0056$.}
    \label{fig:Prewetprofiles}
\end{figure}

Note that path $A$ represents a ``prewetting" path because the adsorption diverges upon reaching bulk coexistence (wet state, cf. Fig.~\ref{fig:trapconfig}c)). In contrast, path $B$ is not a prewetting path because the adsorption remains finite at bulk coexistence (nonwet state, cf. Fig.~\ref{fig:trapconfig}b). The inset in Fig.~\ref{fig:PrewetInterm} corresponds to the computed wave function profiles in a prewet state on path $A$, at $\mu_3/\bar \mu_3 = 0.97$. The wave function profiles are displayed in full in Fig.~\ref{fig:Prewetprofiles}. Notwithstanding their similarity, the wave function profiles in Fig.~\ref{fig:nonwetandwetinterface}b and Fig.~\ref{fig:Prewetprofiles} represent two physically distinct states. The former is a nonwet state at three-phase coexistence, whereas the latter is a prewet state off of three-phase coexistence.
\section{Validity of the theory and outlook to experiments}
Here we ask to what extent the GP mean-field theory at $T=0$ is applicable to the three-component BEC mixtures under consideration. The issues of quantum fluctuations and finite temperature corrections are addressed. The possible impact of van der Waals dispersion forces, neglected in GP theory, is estimated. Furthermore, we ask which atomic species and which trap configurations would be suitable for studying the wetting phenomena experimentally. 

The essential requirement for the validity of the GP theory is the dilute gas limit \cite{Dalfovo,Pita}, in which the particles are only weakly interacting, and which is implemented by assuming  $n_i a_{ii}^3 \ll 1$ for all three components. The dilute gas limit, combined with ultralow temperature ($T=0$ is taken in the calculations), implies that the interactions between atoms are strictly of short and finite range and well represented by Fermi pseudo-potentials of delta-function form. This is very different from the situation in strongly-interacting dense superfluids such as low-$T$ liquid helium, governed by van der Waals forces.

In this theoretical study a large volume and an arbitrarily large number of particles are assumed (cf. grand canonical ensemble in the thermodynamic limit at fixed chemical potentials) so that finite-size effects can be ignored, and finite-size corrections \cite{Dalfovo} neglected, to a first approximation. The confining magnetic trap potential is assumed to be very broad in at least two directions (disk-like confinement), with characteristic harmonic-oscillator lengths of on the order of 10 $\mu$m or more, so that the external potential can be taken to be constant in the calculations. 

Nowadays, especially in the context of multicomponent BEC, the issue of correlation effects beyond GP theory and quantum fluctuations in dilute ultracold Bose gases is receiving much attention, and must be discussed. While correlation effects in the bulk are typically weak (e.g., the depletion of the condensate in the ground state is small compared to that in superfluid helium) \cite{Dalfovo}, there are circumstances in which quantum fluctuations can strongly affect the stability of ``droplets" in multicomponent mixtures. Petrov \cite{Petrov} has shown that correlation effects, taken into account by the Lee-Huang-Yang (LHY) corrections to the energy of a two-component Bose mixture, can stabilize a droplet against collapse when an {\it attractive} interspecies interaction, $G_{ij} < 0$, competes with repulsive intraspecies interactions, $G_{ii}, G_{jj} > 0$, so that the net mean-field energy is much smaller in magnitude than the correlation energy. This effect has been studied recently, for example, in \cite{Edler}. 

In order for the LHY corrections to be important in multicomponent BEC mixtures, it is required that at least one of the interspecies couplings satisfies $K_{ij} \approx - 1$. The fully phase-segregrated components considered in the present work, however, are characterized by repulsive interactions alone, and with $K_{ij} \geq 1$ for all $i < j$. Consequently, neglecting effects of quantum fluctuations on the wetting phase diagrams is justified, to a first approximation. 

At nonzero absolute temperature $T$ one expects two kinds of corrections that may affect the wetting phase diagram. Firstly, the equilibrium interfacial tension $\gamma_{12}(T)$ will be somewhat lowered from $\gamma_{12}(T=0)$ due to thermal fluctuations that excite capillary waves or ``ripplons" on the interface \cite{Mazets}. These excitations  are interfacial Nambu-Goldstone modes associated
with the broken symmetries in this spatially inhomogeneous system. These excitations represent an energy cost reflected in the dispersion relation $\omega \propto k^{3/2}$ (\cite{Mazets}, see also \cite{dyn} and references therein) and an entropy gain, resulting in a net lowering of the interface excess free energy per unit area for $T>0$. The relative correction $(\gamma_{12}(T=0)-\gamma_{12}(T))/\gamma_{12}(T=0)$ is of order $\sqrt{n_ia_{ii}^3}$ \cite{VS} and is negligible for dilute gases. However, in the weak segregation limit, $K_{12}\downarrow 1$, a divergent numerical prefactor was found \cite{VS}, which might have an impact. Secondly, interface fluctuations give rise to an entropic (fluctuation-induced) repulsion between two interfaces a distance $L$ apart \cite{BonnRMP}. This force is favorable to wetting and may lead to a modification of the results obtained here. This refinement is outside the scope of this paper. Also ignored in the present treatment is the possible presence of fluctuation-induced Casimir-like forces induced by Goldstone modes in bulk and/or surface modes. Such forces were found to cause thinning of wetting layers of superfluid helium near the $\lambda$ point and also at lower $T$ in the superfluid regime \cite{Chan,Zandi}.  

Lifshitz theory \cite{DLP,Isra} predicts that interatomic van der Waals forces give rise to long-range forces between two parallel interfaces separating three media (as in Fig.~\ref{fig:prewettingtrapconfig}). The interaction energy per unit area, ${\cal W}$, decays algebraically as a function of the interface separation $L$ in the manner,
\begin{equation}
    {\cal W}= -\frac{A}{12\pi \, L^2} 
\end{equation}
where $A$ is the Hamaker constant. For $T=0$ only the dispersion energy contributes to $A$, and $A$ is proportional to $(n_1^2-n_3^2)(n_2^2-n_3^2)$, with $n_i$ the refractive index of medium $i$ in the visible frequency range. 

For a dilute Bose gas and $T \downarrow 0$ a good approximation is the ``classical" result $n^2 = \alpha \rho$, with $\alpha$ the atomic polarizability \cite{Morice}. From polarizability data \cite{PolMolPhys}, and assuming a typical density $\rho = 10^{14}$ cm$^{-3}$ and a separation $L = 10^{-7}$m,  $|{\cal W}|$ turns out to be of order $10^{-23}$Nm$^{-1}$ for two condensate surfaces separated by vacuum ($n=1$). This can be considered to be an upper bound for our purposes. Assuming three condensates of the same isotope (e.g., $^{87}$Rb) and different hyperfine states, and taking data from \cite{XiaWang}, the order of magnitude of $|{\cal W}|$ is $10^{-32}$Nm$^{-1}$, assuming the same density $\rho$ in each condensate. This value can be considered to be a lower bound. 

In order to assess its importance, ${\cal W}$ must be compared with a typical order of magnitude of an interfacial tension $\gamma_{ij} \approx \sqrt{K_{ij}-1} \,P (\xi_i+\xi_j) $. Using $P = 2\pi  \hbar^2 a \rho^2 /m$ and $\xi = 10^{-7}$m, $\gamma $ is of order  $\sqrt{K_{ij}-1}\;10^{-18}$Nm$^{-1}$. Based on these order of magnitude estimates, the effect of long-range forces, neglected in GP theory, is expected to be unimportant, except possibly in the weak-segregation limit $K_{ij} \downarrow 1$. 

We now comment on the possible choices of atomic species to be used in experiment for studying wetting in phase-segregated BEC mixtures. In \cite{JS}, and independently \cite{MuellerPC,KetterlePC}, it was suggested that  three different hyperfine states of $^{87}$Rb be employed. For that option to satisfy the three immiscibility conditions $K_{ij} > 1$, for all $i<j$, one would have to tune one interspecies scattering length using Feshbach resonance. Other options that were suggested, without prejudice as to whether they satisfy the immiscibility conditions, are the three lowest hyperfine states $|F=1, m_F = 1,0,-1 \rangle$ of $^6$Li, or the three bosonic isotopes (of spin 0) $^{172}$Yb,$^{174}$Yb and $^{176}$Yb \cite{UedaPC}. 

Note that one must be aware of complications that may arise in case the intraspecies scattering lengths $a_{ii}$ differ much between species. For example, in a two-component mixture with $a_{11}>0$ much larger than $a_{22}>0$, in a harmonic trap, or under gravity, the enhanced intraspecies repulsion in component 2 may induce an {\it apparent} demixing even when $K_{12} < 1$, i.e., even when the mixed phase is the equilibrium state. This effect is called ``buoyancy" and it can be confused with phase segregation. Even a flat-bottom box trap requires gravity compensation to avoid buoyancy, but this compensation can only be done for one component \cite{KetterlePC}.

The following set-up has been suggested for an atom trap which may permit the observation of wetting states depicted in Fig.~\ref{fig:trapconfig} in a phase-segregated three-component BEC.
A one-dimensional vertical optical lattice of disk-like traps can be used. Each horizontal ($x-y$) trap width is about 10 to 20 $\mu$m, corresponding to a harmonic oscillator frequency of about 100 Hz. Each vertical ($z$) width is about 100 nm, corresponding to a frequency of about 10 kHz. In this set-up gravity is not an issue~\cite{KetterlePC}. Methods for visualizing a BEC and different BEC components in a trap are available and well-established. Direct non-destructive spatial observation of a BEC and a normal phase was achieved using dispersive light scattering~\cite{Andrews}. Spinor BEC phases with different transverse magnetization components were spatially resolved using optical birefringence and microwave transitions~\cite{StamperKurn}. 
 
\section{Conclusion and outlook}
In this work the wetting phase diagram originally derived for a two-component phase-segregated BEC adsorbed at an optical wall is given a new unequivocal realization in a three-component phase-segregated BEC setting without wall boundary condition. The comparison between theory and experiment is more straightforward in this setting because the control parameters, being the s-wave atomic scattering lengths, are directly accessible and some of them tunable, experimentally. In contrast, wall parameters of previous theory are only indirectly accessible in experiment.

A rich diversity of interfacial phase transitions, including degenerate first-order wetting, first-order wetting, prewetting and, notably, critical wetting, are realized in the three-component GP theory without walls. Critical wetting is of outstanding interest, because {\bf i}) experimental observation of critical wetting in classical liquid mixtures has been a veritable challenge \cite{Ragil,RBM}, and {\bf ii}) theoretically, critical wetting features subtle singularities in the surface excess quantities \cite{Parry}. 

Beyond the scope of our present study, but within reach of future investigation, are non-universal critical exponents, which vary continuously with a ratio of lengths, predicted for classical fluid mixtures in density-functional theory \cite{Hauge} that bears similarities to GP theory. In vector models of magnets this ratio depends on the anisotropy \cite{Walden}. In type-I superconductors the length ratio is that of the magnetic penetration depth and the superconducting coherence length \cite{vLH}. Here, in quantum Bose gas mixtures, the pertinent length ratios involve healing lengths and penetration depths in a manner that is yet to be uncovered.  

Acknowledgements. J.I. gratefully acknowledges the plentiful hospitality of Hanoi Pedagogical University 2, and a sabbatical bench fee (K802422N) from the Research Foundation-Flanders (FWO). We thank the Vietnam National Foundation for Science and Technology Development (NAFOSTED) for support under grant Nr 103.01-2023.12. J.I. benefitted from discussions with Wolfgang Ketterle, Mehran Kardar, Erich M\"uller, Masahito Ueda, Jacques Tempere and Bert Van Schaeybroeck.

{}

\begin{thebibliography}{}
\bibitem{Dalfovo} F. Dalfovo, S. Giorgini, L.P. Pitaevskii, and Sandro Stringari, Theory of Bose-Einstein condensation in trapped gases, Rev. Mod. Phys. {\bf 71}, 463 (1999).
\bibitem{Pita} L.P. Pitaevskii and S. Stringari, {\it Bose-Einstein Condensation} (Clarendon, Oxford, 2003).
\bibitem{Inouye} S. Inouye, M.R. Andrews, J. Stenger, H.-J. Miesner, D.M. Stamper-Kurn, and W. Ketterle, Observation of Feshbach resonances in a Bose–Einstein condensate, Nature {\bf 392}, 151 (1998).
\bibitem{Stan} C.A. Stan, M.W. Zwierlein, C.H. Schunck, S.M.F. Raupach, and W. Ketterle, Observation of Feshbach resonances between two different atomic species, Phys. Rev. Lett. {\bf 93}, 143001 (2004).
\bibitem{Chin} C. Chin, R. Grimm, P. Julienne, and E. Tiesinga, Feshbach resonances in ultracold gases, Rev. Mod. Phys. {\bf 82}, 1225 (2010).
\bibitem{Bloch} I. Bloch, J. Dalibard, W. Zwerger, Many-body physics with ultracold gases, Rev. Mod. Phys. {\bf 80}, 885 (2008).
\bibitem{Gaunt} A.L. Gaunt, T.F. Schmidutz, I. Gotlibovych, R.P. Smith, and Z. Hadzibabic, Bose-Einstein condensation of atoms in a uniform potential, Phys. Rev. Lett. {\bf 110}, 200406 (2013).
\bibitem{Navon} N. Navon, R.P. Smith, Z. Hadzibabic, Quantum gases in optical boxes, Nature Physics {\bf 17}, 1334 (2021).
\bibitem{deG} P.-G. de Gennes, Wetting: statics and dynamics, Rev. Mod. Phys. {\bf 57}, 827 (1985).
\bibitem{Cahn} J.W. Cahn, Critical point wetting, J. Chem. Phys. {\bf 66}, 3667 (1977).
\bibitem{ES} C. Ebner and W. F. Saam, New phase-transition phenomena in thin argon films, Phys. Rev. Lett. {\bf 38}, 1486 (1977).
\bibitem{MC} M.R. Moldover and J.W. Cahn, An interface phase transition: complete to partial wetting, Science {\bf 207}, 1073 (1980).
\bibitem{Dietrich} S. Dietrich, Wetting phenomena, in {\it Phase Transitions and Critical Phenomena}, edited by C. Domb and J.L. Lebowitz (Academic, London), Vol. 12, 1 (1988).
\bibitem{BonnRMP} D. Bonn, J. Eggers, J. Indekeu, J. Meunier and E. Rolley, Wetting and spreading, Rev. Mod. Phys. {\bf 81}, 739 (2009).
\bibitem{BonnRoss} D. Bonn and D. Ross, Wetting transitions, Rep. Prog. Phys. {\bf 64}, 1085 (2001).
\bibitem{Ragil} K. Ragil, J. Meunier, D. Broseta, J.O. Indekeu, and D. Bonn, Experimental observation of critical wetting, Phys. Rev. Lett. {\bf 77}, 1532 (1996).
\bibitem{RBM} D. Ross, D. Bonn and J. Meunier, Observation of short-range critical wetting, Nature {\bf 400}, 737 (1999).
\bibitem{Kozhev} V.F. Kozhevnikov, M.J. Van Bael, P.K. Sahoo, K. Temst, C. Van Haesendonck, A.Vantomme and J.O. Indekeu, Observation of wetting-like phase transitions in a surface-enhanced type-I superconductor, New J. Phys. {\bf 9}, 75 (2007).
\bibitem{IvL} J.O. Indekeu and J.M.J. van Leeuwen, Interface delocalization transition in type-I superconductors, Phys. Rev. Lett. {\bf 75}, 1618 (1995).
\bibitem{IVS} J.O. Indekeu and B. Van Schaeybroeck, Extraordinary wetting phase diagram for mixtures of Bose-Einstein condensates, Phys. Rev. Lett. {\bf 93}, 210402 (2004).
\bibitem{VSIc} B. Van Schaeybroeck and J.O. Indekeu, Critical wetting, first-order wetting, and prewetting phase transitions in binary mixtures of Bose-Einstein condensates, Phys. Rev. A {\bf 91}, 013626 (2015).
\bibitem{Rychtarik} D. Rychtarik, B. Engeser, H.C. Nägerl, and R. Grimm, Two-dimensional Bose-Einstein condensate in an optical surface trap, Phys. Rev. Lett. {\bf  92}, 173003 (2004). 
\bibitem{RW} J.S. Rowlinson, B. Widom, {\it Molecular Theory of Capillarity} (Oxford: Clarendon) (1982).
\bibitem{IK} J.O. Indekeu and K. Koga, Wetting and nonwetting near a tricritical point, Phys. Rev. Lett. {\bf 129}, 224501 (2022).
\bibitem{PR} A.O. Parry and C. Rascon, 
Wetting, algebraic curves and conformal invariance, Phys. Rev. Lett. {\bf 133}, 238001 (2024).
\bibitem{Ho} T.-L. Ho and V.B. Shenoy, Binary Mixtures of Bose Condensates of Alkali Atoms, Phys. Rev. Lett. {\bf 77}, 3276 (1996).
\bibitem{Ao} P. Ao and S.T. Chui,  Binary Bose-Einstein condensate mixtures in weakly and strongly segregated phases, Phys. Rev. A {\bf 58}, 4836 (1998).
\bibitem{sasaki} K. Sasaki, N. Suzuki, D. Akamatsu, and H. Saito, Rayleigh-Taylor instability and mushroom-pattern formation in a two-component Bose-Einstein condensate, Phys. Rev. A {\bf 80}, 063611 (2009).
\bibitem{sasakiC} K. Sasaki, N. Suzuki, and H. Saito, Capillary instability in a two-component Bose-Einstein condensate, Phys. Rev. A {\bf 83}, 053606 (2011). 
\bibitem{kobyakov2} D. Kobyakov, V. Bychkov, E. Lundh, A. Bezett, V. Akkerman, and M. Marklund, Interface dynamics of a two-component Bose-Einstein condensate driven by an external force, Phys. Rev. A {\bf 83}, 043623 (2011). 
\bibitem{takeuchi} H. Takeuchi and K. Kasamatsu, Nambu-Goldstone modes in segregated Bose-Einstein condensates, Phys. Rev. A {\bf 88}, 043612 (2013).
\bibitem{maity} D.K. Maity,  K. Mukherjee, S.I. Mistakidis, S. Das, P.G. Kevrekidis, S. Majumder and P. Schmelcher, Parametrically excited star-shaped patterns at the interface of binary Bose-Einstein condensates, Phys. Rev. A {\bf 102}, 033320 (2020).
\bibitem{Naidon} P. Naidon, D.S. Petrov, Mixed Bubbles in Bose-Bose Mixtures, Phys. Rev. Lett. {\bf 126}, 115301 (2021). 
\bibitem{ruban} V.P. Ruban, Capillary flotation in a system of two immiscible Bose-Einstein condensates, JETP Lett. {\bf 113}, 814 (2021).
\bibitem{myatt} C.J. Myatt, E.A. Burt, R.W. Ghrist, E.A. Cornell and C.E. Wieman, Production of two overlapping Bose-Einstein condensates by sympathetic cooling, Phys. Rev. Lett. {\bf 78}, 586 (1997).
\bibitem{hall} D.S. Hall, M.R. Matthews, J.R. Ensher, C.E. Wieman and E.A. Cornell, Dynamics of component separation in a binary mixture of Bose-Einstein condensates, Phys. Rev. Lett. {\bf 81}, 1539 (1998).
\bibitem{miesner} H.-J. Miesner, D.M. Stamper-Kurn, J. Stenger, S. Inouye, A.P. Chikkatur and W. Ketterle, Observation of metastable states in spinor Bose-Einstein condensates, Phys. Rev. Lett. {\bf 82}, 2228 (1999). 
\bibitem{stamper2} D.M. Stamper-Kurn, H.-J. Miesner, A.P. Chikkatur, S. Inouye, J. Stenger and W. Ketterle, Quantum tunneling across spin domains in a Bose-Einstein condensate, Phys. Rev. Lett. {\bf 83}, 661 (1999).
\bibitem{matthews} M.R. Matthews, B.P. Anderson, P.C. Haljan, D.S. Hall, C.E. Wieman and E.A. Cornell, Vortices in a Bose-Einstein condensate, Phys. Rev. Lett. {\bf 83}, 2498 (1999).
\bibitem{modugno} G. Modugno, M. Modugno, F. Riboli, G. Roati and M. Inguscio, Two atomic species superfluid, Phys. Rev. Lett. {\bf 89}, 190404 (2002).
\bibitem{papp2} S. Papp, J. Pino and C. Wieman, Tunable miscibility in a dual-species Bose-Einstein condensate, Phys. Rev.  Lett. {\bf 101}, 040402 (2008).
\bibitem{Thalhammer} G. Thalhammer, G. Barontini, L. De Sarlo, J. Catani, F. Minardi, and M. Inguscio, Double species Bose-Einstein condensate with tunable interspecies interactions, Phys. Rev. Lett. {\bf 100}, 210402 (2008).
\bibitem{tojo} S. Tojo, Y. Taguchi, Y. Masuyama, T. Hayashi, H. Saito, and T. Hirano, Controlling phase separation of binary Bose-Einstein condensates via mixed-spin-channel Feshbach resonance, Phys. Rev. A {\bf 82}, 033609 (2010).
\bibitem{mccarron} D.J. McCarron, H.W. Cho, D.L. Jenkin, M.P. Köppinger and S.L. Cornish, Dual-species Bose-Einstein condensate of Rb 87 and Cs 133, Phys. Rev. A {\bf 84}, 011603(R) (2011).
\bibitem{wang} F. Wang, X. Li, D. Xiong, D. Wang,  A double species 23Na and 87Rb Bose–Einstein condensate with tunable miscibility via an interspecies Feshbach resonance, J. Phys. B: At. Mol. Opt. Phys. {\bf 49}, 015302 (2015).
\bibitem{Wacker} L. Wacker, N.B. J\o{}rgensen, D. Birkmose, R. Horchani, W. Ertmer, C. Klempt, N. Winter, J. Sherson, and J.J. Arlt, Tunable dual-species Bose-Einstein condensates of 39K and 87Rb, Phys. Rev. A {\bf 92}, 053602 (2015).
\bibitem{Grobner} M. Gr\"obner, P. Weinmann, E. Kirilov, H.-C. N\"{a}gerl, P.S. Julienne, C. Ruth Le Sueur, and J.M. Hutson, Observation of interspecies Feshbach resonances in an ultracold 39K-133Cs mixture and refinement of interaction potentials, Phys. Rev. A {\bf 95}, 022715 (2017).
\bibitem{burchianti} A. Burchianti, C. D'Errico, S. Rosi, A. Simoni, M. Modugno, C. Fort and F. Minardi,  Dual-species Bose-Einstein condensate of K 41 and Rb 87 in a hybrid trap, Phys. Rev. A {\bf 98}, 063616 (2018). 
\bibitem{lee} K.L. Lee, N.B. Jørgensen, L.J. Wacker, M.G. Skou, K.T. Skalmstang, J.J. Arlt and N.P. Proukakis,  Time-of-flight expansion of binary Bose–Einstein condensates at finite temperature, New J. Phys. {\bf 20}, 053004 (2018).
\bibitem{Kwon} K. Kwon, K. Mukherjee, S. Huh, K. Kim, S.I. Mistakidis, D.K. Maity, P.G. Kevrekidis, S. Majumder, P. Schmelcher, J.-y. Choi, Spontaneous formation of star-shaped surface patterns in a driven Bose-Einstein condensate, Phys. Rev. Lett. {\bf 127}, 113001 (2021).
\bibitem{wilson} K.E. Wilson, A. Guttridge, J. Segal and S.L. Cornish, Quantum degenerate mixtures of Cs and Yb, Phys. Rev. A, {\bf 103}, 033306 (2021).
\bibitem{MaPe} Y. Ma, C. Peng and X. Cui, Borromean droplet in three-component ultracold Bose Gases, Phys. Rev. Lett. {\bf 127}, 043002 (2021).
\bibitem{Blom} E. Blomquist, A. Syrwid and E. Babaev, Borromean supercounterfluidity, Phys. Rev. Lett. {\bf 127} 255303 (2021).
\bibitem{Keiler} K. Keiler, S.I. Mistakidis, and P. Schmelcher, Polarons and their induced interactions in highly imbalanced triple mixtures, Phys. Rev. A {\bf 104}, L031301 (2021).
\bibitem{JS} K. Jimbo and H. Saito, Surfactant behavior in three-component Bose-Einstein condensates, Phys. Rev. A {\bf 103}, 063323 (2021).
\bibitem{Edler} D. Edler, L.A. Pena Ardila, C.R. Cabrera and L. Santos, Anomalous buoyancy of quantum bubbles in immiscible Bose mixtures, Phys. Rev. Res. {\bf 4}, 033017 (2022).
\bibitem{Saboo} A. Saboo, S. Halder, S. Das and S. Majumder, Rayleigh-Taylor instability in a phase-separated three-component Bose-Einstein condensate,  Phys. Rev. A {\bf 108}, 013320 (2023). 	
\bibitem{Roberts} D.C. Roberts and M. Ueda, Stability analysis for $n$-component Bose-Einstein condensate, Phys. Rev. A {\bf 73}, 053611 (2006).
\bibitem{Timmermans} E. Timmermans, Phase Separation of Bose-Einstein Condensates, Phys. Rev. Lett. {\bf 81}, 5718 (1998).
\bibitem{Bar} R.A. Barankov, Boundary of two mixed Bose-Einstein condensates, Phys. Rev. A {\bf 66}, 013612 (2002).
\bibitem{VS} B. Van Schaeybroeck, Interface tension of Bose-Einstein condensates, Phys. Rev. A  {\bf 78}, 023624 (2008). See also the Addendum, B. Van Schaeybroeck, Phys. Rev. A {\bf 80}, 065601 (2009).
\bibitem{Istatic} J.O. Indekeu, C.-Y. Lin, T.V. Nguyen, B. Van Schaeybroeck and T.H. Phat, Static interfacial properties of Bose-Einstein-condensate mixtures, Phys. Rev. A {\bf 91}, 033615 (2015).
\bibitem{Parry} For insight in the subtlety of short-range critical wetting, see A.O. Parry and C. Rascon, The trouble with critical wetting, J. Low Temp. Phys. {\bf 157}, 149 (2009).
\bibitem{VSNI} B. Van Schaeybroeck, P. Navez, and J. O. Indekeu, Interface potential and line tension for Bose-Einstein condensate mixtures near a hard wall, Phys. Rev. A {\bf 105}, 053309 (2022).
\bibitem{SJdG} D. Saint-James and P.-G. de Gennes, Phys. Lett. {\bf 7}, 306 (1963), see also P.-G. de Gennes, Superconductivity of Metals and
Alloys (Addison-Wesley, Reading, 1966).
\bibitem{IvL2} J.O. Indekeu and J.M.J. van Leeuwen, Wetting, prewetting and surface transitions in type-I superconductors, Physica C {\bf 251}, 290 (1995).
\bibitem{Petrov} D.S. Petrov, Quantum Mechanical Stabilization of a Collapsing Bose-Bose Mixture, Phys. Rev. Lett. {\bf 115}, 155302 (2015).
\bibitem{Mazets} I.E. Mazets, Waves on an interface between two phase-separated Bose-Einstein condensates, Phys. Rev. A {\bf 65}, 033618 (2002).
\bibitem{dyn} J.O. Indekeu, N.V. Thu, C. Lin, and T.H. Phat, Capillary-wave dynamics and interface structure modulation in binary Bose-Einstein condensate mixtures, Phys. Rev. A {\bf 97}, 043605 (2018).
\bibitem{Chan} R. Garcia and M.H.W. Chan, Phys. Rev. Lett. {\bf 83}, 1187 (1999).
\bibitem{Zandi} R. Zandi, J. Rudnick, and M. Kardar, Phys. Rev. Lett. {\bf 93}, 155302 (2004).
\bibitem{DLP} I.E. Dzyaloshinskii, E.M. Lifshitz, and L.P. Pitaevskii, The general theory of van der Waals forces, Adv. Phys. {\bf 10}, 165 (1961).
\bibitem{Isra} J. Israelachvili, {\it Intermolecular and surface forces}, 2nd ed. (Academic Press, Amsterdam, 1991)
\bibitem{Morice} O. Morice, Y. Castin, and J. Dalibard, Refractive index of a dilute Bose gas, Phys. Rev. A {\bf 51}, 3896 (1995). 
\bibitem{PolMolPhys} P. Schwerdtfeger and J.K. Nagle, 2018 Table of static dipole polarizabilities of the neutral elements in the periodic table, Mol. Phys. {\bf 117}, 1200 (2019).
\bibitem{XiaWang} X. Wang, J. Jiang, L.-Y. Xie, D.-H. Zhang, and C.-Z. Dong, Polarizabilities and tune-out wavelengths of the hyperfine ground states of $^{87}$,$^{85}$Rb, Phys. Rev. A {\bf 94}, 052510 (2016).
\bibitem{MuellerPC} Erich M\"uller, private discussion (2024).
\bibitem{KetterlePC} Wolfgang Ketterle, private discussion (2024).
\bibitem{UedaPC} Masahito Ueda, private discussion (2024).
\bibitem{Andrews} M.R. Andrews, M.-O. Mewes, N.J. van Druten, D.S. Durfee, D. M. Kurn, and W. Ketterle, Direct, non-destructive observation of a Bose condensate, Science {\it 273}, 84 (1996).
\bibitem{StamperKurn} D.M. Stamper-Kurn, Seeing Spin Dynamics in Atomic Gases, in ``From Atomic to Mesoscale", p. 61, World Scientific (2015).
\bibitem{Hauge} E.H. Hauge, Landau theory of wetting in systems with a two-component order parameter, Phys. Rev. B {\bf 33}, 3322 (1986).
\bibitem{Walden} C.J. Walden, B.L. Gy\"orffy, and A.O. Parry, Nonuniversal anisotropy dependence of critical-wetting exponents in a vector model, Phys. Rev. B {\bf 42}, 798 (1990).
\bibitem{vLH} J.M.J. van Leeuwen  and E.H. Hauge, The effective interface potential for a superconducting layer, J. Stat. Phys. {\bf 87}, 1335 (1997).
\end{thebibliography}
\end{document}